\documentclass[12pt,a4paper]{article}
\usepackage{graphicx}% Include figure files
\usepackage{dcolumn}% Align table columns on decimal point
\usepackage{bm}% bold math
\usepackage{amsmath,amsfonts,amssymb,mathrsfs}
\usepackage{slashed}
\usepackage{braket,xcolor}
\usepackage{verbatim}
\usepackage{subcaption}
\usepackage{multirow}
\usepackage{amsfonts,mathtools,resizegather}
\usepackage[utf8]{inputenc}
\usepackage{caption,jheppub}
\DeclareMathOperator{\sech}{sech}

\title{\boldmath {Krylov complexity in saddle-dominated scrambling}}

\author[a]{Budhaditya Bhattacharjee,}
\author[b]{Xiangyu Cao,}
\author[a]{Pratik Nandy,}
\author[a]{and Tanay Pathak}

\affiliation[a]{Centre for High Energy Physics, Indian Institute of Science,\\C.V. Raman Avenue, Bangalore 560012, India}
\affiliation[b]{Laboratoire de Physique de l'Ecole Normale Sup\'erieure, ENS, Universit\'e PSL,\\
CNRS, Sorbonne Universit\'e, Universit\'e de Paris, 75005 Paris, France}

\emailAdd{budhadityab@iisc.ac.in}
\emailAdd{xiangyu.cao@ens.fr}
\emailAdd{pratiknandy@iisc.ac.in}
\emailAdd{tanaypathak@iisc.ac.in}

\abstract{In semi-classical systems, the exponential growth of the out-of-time-order correlator (OTOC) is believed to be the hallmark of quantum chaos. However, on several occasions, it has been argued that, even in integrable systems, OTOC can grow exponentially due to the presence of unstable saddle points in the phase space. In this work, we probe such an integrable system exhibiting saddle-dominated scrambling through Krylov complexity and the associated Lanczos coefficients. In the realm of the universal operator growth hypothesis, we demonstrate that the Lanczos coefficients follow the linear growth, which ensures the exponential behavior of Krylov complexity at early times. The linear growth arises entirely due to the saddle, which dominates other phase-space points even away from itself. Our results reveal that the exponential growth of Krylov complexity can be observed in integrable systems with saddle-dominated scrambling and thus need not be associated with the presence of chaos.}

\begin{document} 
\maketitle
\flushbottom

\section{Introduction}
\label{sec:intro}

In the past few years, the study of chaotic systems has garnered much attention. While a proper understanding of quantum chaos still eludes us, much progress has been made in this direction. The fundamental problem is, of course, the interpretation of the definition of quantum chaos. The phenomena of chaos are well-defined classically, namely the exponential deviation of the phase-space trajectories for small initial perturbations. Such sensitivity is usually termed as the ``butterfly effect" \cite{Roberts:2014ifa, Roberts:2016wdl, Cotler:2017myn}. However, in quantum mechanics, one runs into trouble as the definition of trajectories become ill-defined. One resorts to various probes and measures in such cases. To date, the most versatile and extensively used probe is the level-statistics of the quantum mechanical Hamiltonian \cite{PhysRevE.81.036206, DAlessio:2015qtq}. Chaotic systems tend to follow the Wigner-Dyson statistics, while the integrable ones follow the statistics of Poisson distribution. Various other measures have been studied, especially in conformal field theories where entanglement plays a pivotal role \cite{Kudler-Flam:2019kxq}.

However, an indirect but more convenient way to probe the integrable and the chaotic nature of the Hamiltonian is to study the operator growth in individual cases. In chaotic systems, operators are supposed to grow more rapidly than their integrable counterparts. The most sought-after measure that has been used is the out-of-time-ordered-correlator (OTOC) \cite{Rozenbaum:2016mmv, Hashimoto:2017oit, Nahum:2017yvy, 2017, Shen:2017kez, Khemani:2017nda, PhysRevB.100.035112,  Pilatowsky-Cameo:2019qxt, Xu:2019lhc,  Rozenbaum:2019nwn, Hashimoto:2020xfr, Styliaris:2020tde, Akutagawa:2020qbj,  Zhou:2021syv, Xu:2022vko}. It uses a probe operator to detect the overlap from the time evolution of the reference operator. It is defined as a squared commutator with a thermal expectation value of inverse temperature $\beta$ and given by
\begin{align}
    \mathrm{OTOC}(t) = -\braket{[V(0), \mathcal{O}(t)]^2}_{\beta}\,.
\end{align}
Here $\mathcal{O}(t)$ is the given operator that evolves over time, and the overlap is detected by a constant probe operator $V(0)$. In other words, it can also be understood as the overlap between two states, which has been operated with different ordering of $V(0)$ and $\mathcal{O}(t)$. However, the above definition suffers a regularization problem, especially when it is extended to the continuum version, i.e., in quantum field theory. To resolve this, one uses the smearing of two separate commutators using the thermal density matrix $\rho = \exp(-\beta H)/Z$ as \cite{Maldacena:2015waa, Trunin:2020vwy}
\begin{align}
    \mathrm{OTOC}(t) = -\braket{\rho^{1/2}[V(0), \mathcal{O}(t)] \rho^{1/2}[V(0), \mathcal{O}(t)]}_{\beta}\,.
\end{align}
It has been proposed that for a chaotic system, the OTOC grows exponentially with the coefficient of exponent, known as Lyapunov exponent, saturating the well-known chaos bound \cite{Maldacena:2015waa}\footnote{See \cite{Kundu:2021qcx, Kundu:2021mex, Hashimoto:2021afd} for recent developments on the subleading terms and the energy bound.}. However, several recent studies have also exposed the dubious role of OTOC; the exponential growth of OTOC does not always imply the chaotic nature of the given system \cite{Xu:2019lhc, Hashimoto:2020xfr, Rozenbaum:2019nwn,  Pilatowsky-Cameo:2019qxt}.
It can simply happen due to unstable saddle points in the classical phase space. See \cite{Caputa:2016tgt, Liu:2018iki, Hummel:2018brt, Pappalardi:2018frz, Ali:2019zcj, Kelly:2019bvd, Lambert:2021epa, Kidd:2020mtu, Bhattacharyya:2020art} for some related examples. As argued in \cite{Xu:2019lhc}, this phenomenon, known as scrambling, should be distinguished from quantum chaos. The inference is that the exponential growth of OTOC can be possible even in integrable systems, rendering OTOC, in some instances, to be a poor indicator of quantum chaos.

In this work, we turn to another interesting probe, namely the Krylov complexity (K-complexity in short), to study the operator growth in systems possessing such saddles. The K-complexity was first introduced in \cite{Parker:2018yvk} to examine the universal feature of operator growth. The objective is to probe the Heisenberg evolution of some simple initial Hermitian operators. A simple operator can become extremely complicated depending upon the nature of the Hamiltonian, and the given initial operator \cite{Roberts:2014isa}. However, we do not need a probe operator to capture the growth here. Instead, we construct a set of basis, known as Krylov basis, from the complicated nested operators coming from the expansion of Baker-Campbell-Hausdorff expansion of the time-evolved operator. Construction of this Krylov basis set is known as the Lanczos algorithm, and the growth is compactly encoded in a series of coefficients termed Lanczos coefficients. The universal operator growth hypothesis \cite{Parker:2018yvk} posits that the growth of Lanczos coefficients is fastest for a chaotic system. On the other hand, the growth is much slower for integrable and free theories. Interestingly, the validity of this hypothesis goes beyond the semi-classical regime, where OTOC or, more specifically, the Lyapunov exponent is ill-defined. In recent years, the study of operator growth and K-complexity has received significant attention from many-body systems to the conformal field theories and black hole physics \cite{Barbon:2019wsy, Avdoshkin:2019trj, Dymarsky:2019elm, Jian:2020qpp, Rabinovici:2020ryf, Cao:2020zls, Dymarsky:2021bjq, Yates:2021asz, Kar:2021nbm, PhysRevE.104.034112, Caputa:2021sib, Kim:2021okd, Caputa:2021ori,  Patramanis:2021lkx,  Rabinovici:2021qqt,  Trigueros:2021rwj,  Hornedal:2022pkc, Balasubramanian:2022tpr, Fan:2022xaa, Heveling:2022hth}.

\begin{figure}
		\centering
		\begin{subfigure}[b]{0.48\textwidth}
		\centering
		\includegraphics[width=\textwidth]{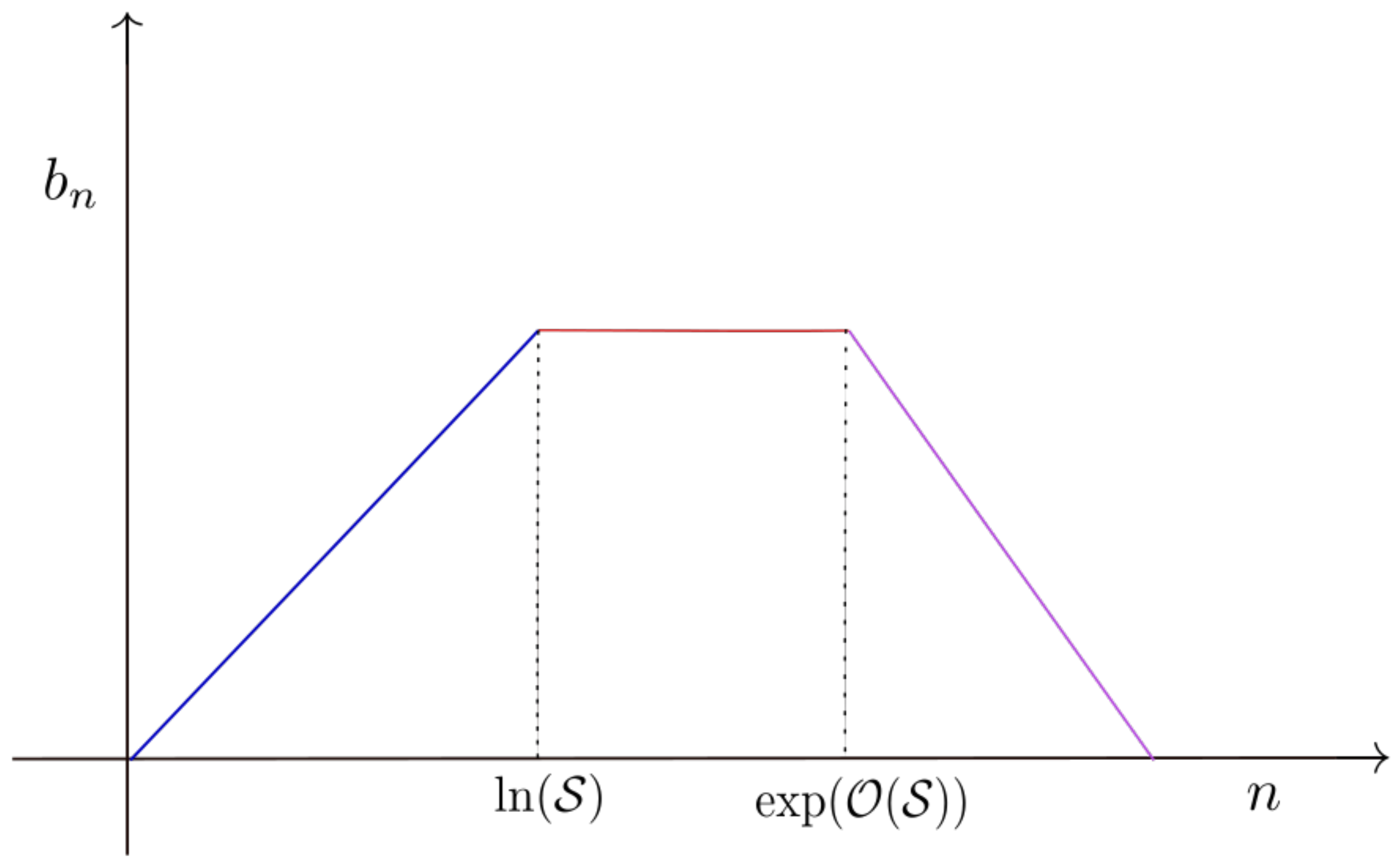}
			\caption{Growth of $b_{n}$ with $n$.}
			\label{fig:bns}
		\end{subfigure}
		\hfill
		\begin{subfigure}[b]{0.48\textwidth}
		\centering
		\includegraphics[width=\textwidth]{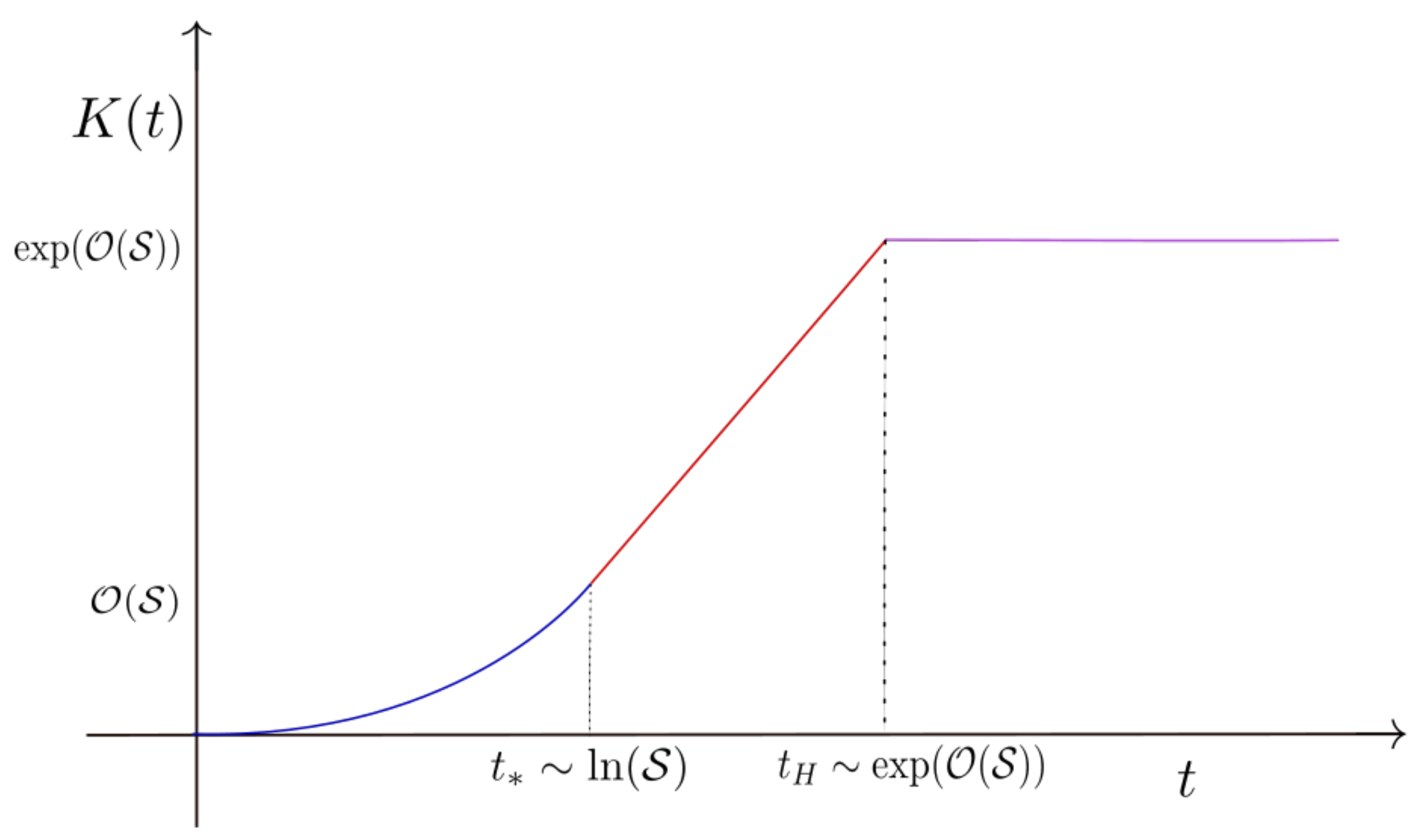}
			\caption{Growth of $K(t)$ with $t$.}
			\label{fig:Kts}
		\end{subfigure}
		\caption{A rough sketch of the behavior of Lanczos coefficients $b_{n}$ with $n$ for a ``chaotic'' system. The linear increase (blue) of Lanczos coefficients corresponds to the exponential growth of K-complexity $K(t)$ up to the time-scale $t_{*} = \ln \mathcal{S}$. After that, $b_n$ saturates to some constant (red), which corresponds to the linear increase of K-complexity. Finally, the decrease of $b_n$ (purple) is the Lanczos descent which marks the saturation of K-complexity.}
		\label{ll}
\end{figure}

The growth of Lanczos coefficients essentially captures the time evolution of the K-complexity (see Fig.\,\ref{ll}). The linear growth of Lanczos coefficients ensures the exponential growth of K-complexity up to the scrambling time $t_{*} \sim \ln \mathcal{S}$, where $\mathcal{S}$ is the entropy of the system,\footnote{We use the notation $\mathcal{S}$ to denote the entropy. This should be differentiated with spin $S$, as we will describe later.} reaching a value of $\mathcal{O}(\mathcal{S})$. This linear growth of $b_n$ is known as \emph{Lanczos ascent}. This is followed by a almost constant growth (post-scrambling) of $b_n$ up to the Heisenberg time $t_{H} \sim \exp(\mathcal{O}(\mathcal{S}))$, where K-complexity saturates a value of $\exp(\mathcal{O}(\mathcal{S}))$. This region is known as \emph{Lanczos plateau}. At very late times, $b_n$ decreases to zero when the whole Krylov space is exhausted, and as a result,  the K-complexity saturates \cite{Barbon:2019wsy, Rabinovici:2020ryf}. Holographically, similar behavior has been conjectured for the volume of the interior of the black hole at late times \cite{Susskind:2014rva, Susskind:2014moa}. 

The main driving force of this work is to study K-complexity in an integrable model exhibiting an unstable fixed point in its phase space. We study a known \textit{integrable} finite-dimensional quantum spin system, known as Lipkin-Meshkov-Glick (LMG) model \cite{LIPKIN1965188, GLICK1965211}  as studied in \cite{Xu:2019lhc}. Despite the model being integrable \cite{Debergh:2001xt, PhysRevE.78.021106, LermaH:2013bax}, OTOC grows exponentially at early times. Hence, OTOC fails to capture the integrability of the model. This drives to check whether K-complexity is better suited in this case. If the universal operator growth hypothesis is strictly restricted to chaotic and integrable systems, then we expect that the Lanczos coefficients might grow sub-linearly, and the K-complexity should follow a power-law growth, in contrast to the exponential growth that holds for the non-integrable case. However, this is not the case for the model under our consideration. In fact, the Lanczos coefficients grow linearly with $n$, and consequently, that shows the exponential growth of K-complexity. In the realm of the universal operator growth hypothesis, this exponential growth is a feature of the chaotic system. In our case, this comes from the saddle-dominated scrambling. Hence, the K-complexity cannot distinguish between the saddle-dominated scrambling in integrable systems and generic chaotic systems. Moreover, the saddle can dominate even in a chaotic system, for example, the Feingold-Peres (FP) model \cite{FEINGOLD1983433, PhysRevA.30.504, PhysRevA.30.509, Xu:2019lhc, Parker:2018yvk}, which we briefly discuss in the Appendix \ref{appendFP}. This suggests that the universal operator growth hypothesis, especially the linear growth of the Lanczos coefficients, must also include the phenomena of saddle-dominated scrambling, with or without the presence of chaos. To the knowledge of the present authors, this fact has not been previously mentioned or examined thoroughly.

As the exponential growth predominantly comes from the saddle, one could wonder if it is possible to devise a microcanonical version of the complexity that can capture the behavior near the saddle. This can be achieved by considering a fixed energy window near the saddle, and we will see that it can be accurately done. Especially, we find the microcanonical K-complexity shows dominant exponential growth near the saddle, which primarily controls the overall behavior of K-complexity. Away from the saddle, the contributions are subdominant but still exponential. This is seen via the linear growth of the Lanczos coefficients away from the saddle. One can note that the Lyapunov exponent away from the saddle is smaller (compared to the one close to the saddle) but still non-zero. Therefore, the microcanonical K-complexity can detect the presence of the unstable saddle even if it is evaluated far from the saddle neighborhood. The growth of Lanczos coefficients away from the saddle is also interesting. For large values of $n$, the odd and even coefficients grow distinctly, contrary to the case at the saddle itself, where both odd and even coefficients grow similarly. This oscillation in the Lanczos coefficients can be traced back to the decay to auto-correlation functions at sufficiently late-times. We numerically study the behavior of the auto-correlation function near the saddle as well as away from it and provide an intuitive analytical argument for the same. We find that indeed as expected, the function decays rapidly near the saddle but decays slowly (with oscillatory behavior) away from the saddle, hence giving rise to the oscillations in the Lanczos coefficients.

 The paper is structured as follows. In section \ref{sec2}, we briefly go over the OTOC calculation and elucidate the results. We also conduct a classical analysis of the LMG model to identify the unstable saddle point(s). In section \ref{sec3}, we provide a brief review of the Krylov complexity and its salient features. We then study the behavior of K-complexity in the LMG model and observe the effect of the unstable saddle on the K-complexity. In section \ref{sec4}, we examine the microcanonical version of the same and observe the impact of the unstable saddle on it. A classical analysis for the poles of the auto-correlation function is also provided, with an emphasis on the explanation of the observed behavior of the K-complexity. We conclude the paper with a brief discussion and outlook in section \ref{conclusion}. In Appendix \ref{appendFP}, we provide a classical analysis of the Feingold-Peres (FP) model (which shows saddle-dominated scrambling despite being chaotic) and study the behavior of K-complexity.

\section{OTOC and saddle-dominated scrambling} \label{sec2}
As our first pass, we attempt to study the quantum Lipkin-Meshkov-Glick (LMG) model \cite{LIPKIN1965188, GLICK1965211} through OTOC. However, before that, we will briefly revisit the classical LMG model, especially the scrambling behavior near the saddle point in the phase space. 

\subsection{Saddles of the LMG model: a classical analysis}
\label{sub:sec:1}
In this section, we study the LMG model classically and infer its scrambling behavior around an unstable saddle point.
The system is described by the Hamiltonian
\begin{equation} \label{LMGclassHam}
    H = x +  J z^{2}\,,
\end{equation}
where the classical variable $x,y,z$ are constrained on a sphere $x^{2} + y^{2} + z^{2} = 1$, and they follow the classical $SU(2)$ algebra $\{x,y\} = z$ and similarly for other variables in cyclic order. Here $\{~,~\}$ is the Poisson bracket. To visualize the phase-space trajectory, we need to solve the Hamilton's equation of motion
\begin{equation}
\frac{\mathrm{d} X_{i}}{\mathrm{d} t} = \{X_{i}, H\}\,, \label{eqnom}
\end{equation}
where $X_i = \{x,y,z\}$ are the phase-space coordinates. Using the above Hamiltonian \eqref{LMGclassHam}, the equations of motion can be explicitly written as
\begin{align}
\frac{\mathrm{d} x}{\mathrm{d} t} = - 2 J y z, ~~~~~ 
\frac{\mathrm{d} y}{\mathrm{d} t} = - z + 2 J x z,~~~~~
\frac{\mathrm{d} z}{\mathrm{d} t} =  y\,. \label{diffeqn3}
\end{align}
To obtain these equations, we have used the following properties of the Poisson bracket
\begin{align}
\{f, g\} &= -\{g, f\}, \;\;~~~~~~~~~~~~~~~~~ \text{(anti-commutativity)}\\
\{f g, h\} &= \{f, h\}g + f\{ g, h\}. \;\; ~~~~~\text{(Leibniz's rule)}
\end{align}
Now, we proceed to detect the saddle points. A saddle point is defined as the point $(x_{0},y_{0},z_{0})$ where $\mathrm{d} X_{i}/\mathrm{d} t = 0 , \; i = 1,2,3$. This gives us the following set of conditions
\begin{align}
    -2 J y z = 0\,,\;\;\;\;-z (1-2 J x) = 0\,,\;\;\;\;y = 0\,.
\end{align}
Clearly, the solution to these are the following
\begin{align}
    y &= 0\,,\;\;\;\; z = 0\,, \;\;\;\;\;\;\;\;\;\;\;\; x = \pm 1\,, \\
    y &= 0\,,\;\;\;\; z = \pm \sqrt{1 - \frac{1}{4 J^{2}}}\,, \;\;\;\; x = \frac{1}{2 J} \,.
\end{align}
It is evident that a saddle point does not exist for $J < 1/2$. To determine whether the saddle is stable or unstable, we need to observe the Jacobian of transformation between the ``coordinates'' $\mathrm{d}x/\mathrm{d}t, \mathrm{d}y/\mathrm{d}t, \mathrm{d}z/\mathrm{d}t$ and $x,y,z$. The matrix is given by
\begin{equation}
\mathcal{J} = \begin{bmatrix}
    0 & -2 J z & -2 J y\\
    2 J z & 0 & 2 J x - 1\\
    0 & 1 & 0
    \end{bmatrix}.
\end{equation}
Now, the eigenvalues at the four saddle points are the following
\begin{align}
    e &= (-\sqrt{2 J - 1}, 0, \sqrt{2 J - 1})\,, ~~~~~~~~~\, (x,y,z) = (1,0,0)\,, \label{eq1}\\
    e &= (0,-i\sqrt{2 J + 1}, i\sqrt{2 J + 1})\,,~~~~~~~\, (x,y,z) = (-1,0,0)\,, \\
    e &= (0, -i\sqrt{4 J^{2}-1} , i\sqrt{4 J^{2}-1})\,,~~~~~ (x,y,z) = \left(\frac{1}{2 J},0,\pm \sqrt{1 - \frac{1}{4 J^{2}}}\right)\,.
\end{align}
We see that the unstable saddle is at the point $(x,y,z) = (1,0,0)$. \par 
At this saddle, the equations of motion linearize. While it can be seen from the linearized equations of motion as well, it is easier to simply observe the result in \eqref{eq1}, which tells us that the normal modes at $(1,0,0)$ will behave as $e^{\lambda t}$ (which is the generic behavior for unstable saddle points) with $\lambda  = \sqrt{2 J -1}$. The eigenvalues indicate the existence of normal mode $a_{\pm}$ that behave as 
\begin{equation}
    \frac{\mathrm{d} a_{\pm}}{\mathrm{d} t} \sim \pm \lambda a_{\pm}\,.
\end{equation}
The solution gives exponential growth with a classical ``Lyapunov'' exponent $\lambda_{L} \sim \lambda =  \sqrt{2 J - 1}$. The presence of this saddle emulates a chaotic operator growth, as demonstrated with OTOCs with a Lyapunov coefficient $\lambda_{\text{OTOC}}$ in \cite{Xu:2019lhc}. Of course, such behavior is not actually chaotic since the actual exponential growth takes place only near the saddle point itself. The scenario, which we numerically study in the next subsection, corresponds to $J = 2$.

\subsection{OTOC and LMG model}

In this subsection, we consider the quantum mechanical version of the LMG model, studied in \cite{Xu:2019lhc}. The quantum LMG Hamiltonian (with $J = 2$)\footnote{We compute K-complexity for various $J$ in later sections and infer the conclusion with analytic computations.} is given by
\begin{equation}\label{Hamquant}
    H = \hat{x} + 2 \hat{z}^{2}\,,
\end{equation}
where we define $\hat{x} = \hat{S}_x/S, \hat{y} = \hat{S}_y/S, \hat{z} = \hat{S}_z/S$, which are the rescaled $SU(2)$ spin operators with spin $S$. They satisfy the commutation relation $[\hat{x},\hat{y}] = i \hbar_{\text{eff}}\hat{z}$ with cyclic order, where $\hbar_{\text{eff}} = 1/S$ is the effective Planck's constant. The classical limit is achieved by taking $\hbar_{\text{eff}} \rightarrow 0$, which is equivalent to the large-$S$ expansion. This fact is previously noted in \cite{Cotler:2017myn, Yin:2020oze}.\par
Following the work of \cite{Xu:2019lhc}, we study the OTOC for this system. The reference operator is chosen to be $\mathcal{O} = \hat{z}$. The OTOC is defined as
\begin{equation}
    \mathrm{OTOC} \,(t) = \frac{1}{\hbar^{2}_{\text{eff}}}\frac{\text{Tr}\Big( [\mathcal{O}(t),\mathcal{O}] [\mathcal{O}(t),\mathcal{O}]^{\dagger} \Big)}{\text{Tr}(\mathbf{1})}\,. \label{otoc}
\end{equation}
Here, the evolution of $\mathcal{O}$ is given by Heisenberg evolution of operators
$\mathcal{O}(t) = e^{i H t/h_{\mathrm{eff}}} \, \mathcal{O} \, e^{-i H t/h_{\mathrm{eff}}}$, and OTOC is supposed to capture the growth of $\mathcal{O}(t)$ as time evolves. It is widely believed that for chaotic Hamiltonians, the growth would be more rapid than the integrable systems. In fact, along this line, the operator growth has been previously studied in \cite{Roberts:2014isa}, in the context of precursor growth. In the holographic picture, the growth has been intuitively understood as a tensor network from the boundary side, which is dual to the Einstein-Rosen bridge aided by localized shocks in bulk. However, in this work, we do not talk about precursor growth and do not consider any holographic dual geometry of the operator growth in the following discussions.

\begin{figure}
    \centering
    \includegraphics[width=10cm, height=6.5cm]{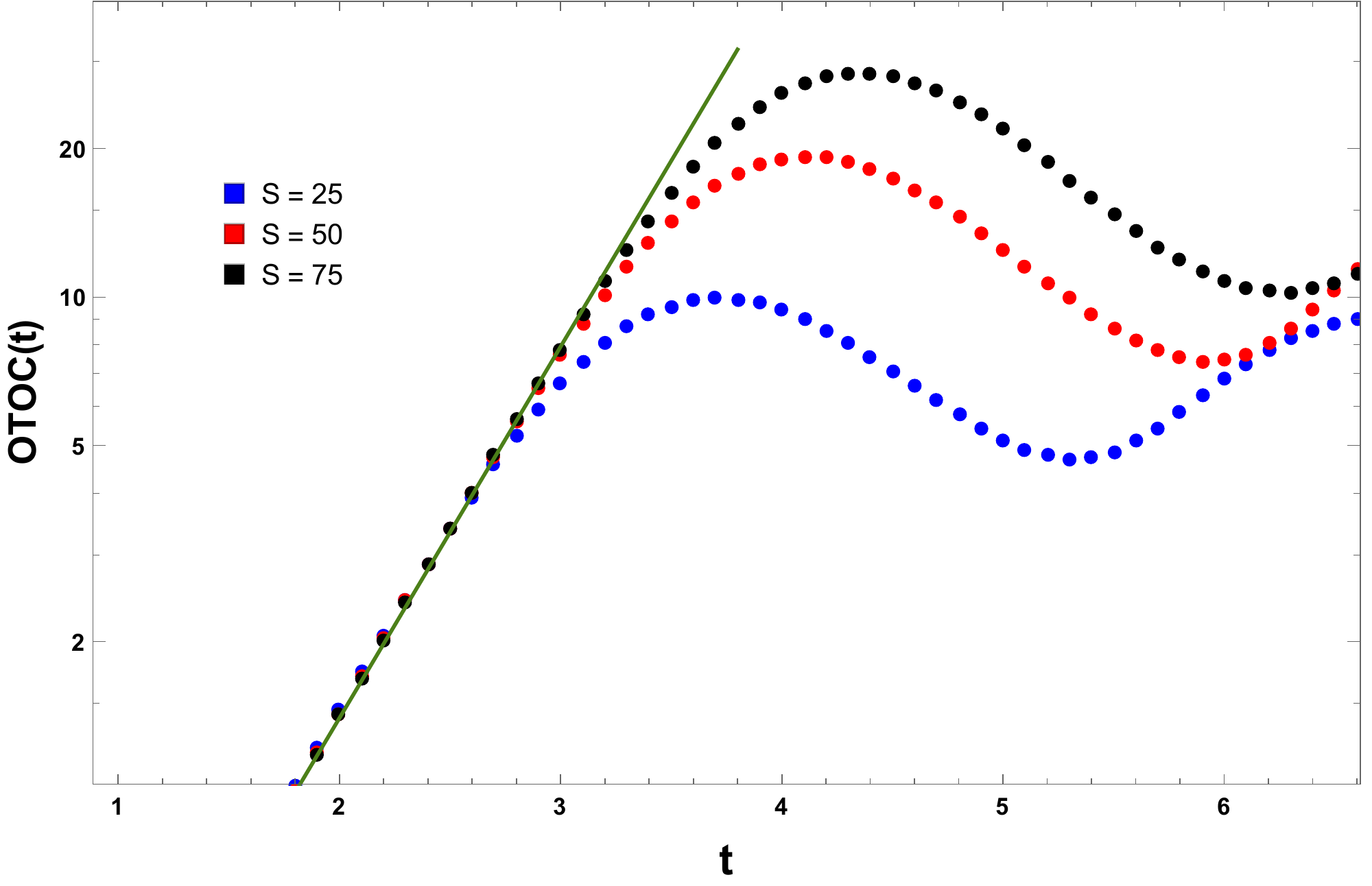}
    \caption{Behavior of OTOC for quantum LMG model (log plot). The OTOC is calculated for spin values $S = 25, 50$ and $75$. The early time behavior of OTOCs are fitted to an exponential function $ \sim e^{\lambda_{\text{OTOC}}t}$. We obtain $\lambda_{\text{OTOC}} = 1.73446 > \lambda_{\text{saddle}} = \sqrt{3}$, which satisfies the classical bound \eqref{cbound} on $\lambda_{\text{OTOC}}$.  }
    \label{fig1}
\end{figure}

As the time evolves, the operator $\mathcal{O}(t)$ becomes extremely complicated. Hence, the OTOC of Eq.\eqref{otoc} cannot be calculated analytically. There exist a few special cases (for example, for simple or inverted harmonic oscillators) where an explicit expression of $\mathcal{O}(t)$ can be obtained for various seed operators \cite{Bhattacharyya:2020art}. However, Eq.\eqref{otoc} can be evaluated numerically. With the initial operator $\hat{z}$, we numerically compute the OTOC for various spins. The result is shown in Fig.\,\ref{fig1}. We see that as we increase the spin, the saturation takes place at a later time. The classical limit is, of course, at $S \rightarrow \infty$. For finite $S$, the Lyapunov exponent obeys the bound \cite{Xu:2019lhc}
\begin{align}
    \lambda_{\text{OTOC}} \geq \lambda_{\text{saddle}}\,. \label{cbound}
\end{align}

where $\lambda_{\mathrm{saddle}}$ can be obtained from the exponential growth near the saddle. The bound is saturated at the classical limit $S \rightarrow \infty$. This bound is different with bound involving $\lambda_{\mathrm{Chaos}}$, which directs the authors of \cite{Xu:2019lhc} to claim that the notion of chaos must be distinguished from scrambling. Even, from the OTOC computations, we observe that there is a period of exponential growth roughly within the time range $1 \leq t \leq t_{\text{Ehrenfest}}$, where the Ehrenfest time (also known as the Scrambling time) $t_{\text{Ehrenfest}}$ is related to the spin $S$ as $t_{\text{Ehrenfest}} \sim \ln(S)$. We can directly compute $\lambda_{\text{OTOC}}$ from Fig.\,\ref{fig1}, by numerical fitting. In fact, we see that the bound \eqref{cbound} is obeyed, i.e., $\lambda_{\text{OTOC}} = 1.73446  > \lambda_{\text{saddle}} \equiv \omega_{\text{saddle}} = \sqrt{3}$.

As we have discussed earlier, the exponential growth of the OTOC at early times is believed to be an indicative measure of chaos. While that is true for a large class of systems, the above result demonstrates that, at least for the integrable LMG model considered above, the OTOC behaves in the same way as it does for a chaotic system. Hence, OTOC fails to capture the integrability of the LMG model. This is entirely due to the existence of saddle, and we see that saddle-dominated scrambling overwhelms the integrable nature of the LMG Hamiltonian.

In the next section, we introduce the K-complexity as a probe to the operator growth, and we calculate the K-complexity in the LMG model. We will see that K-complexity is also sensitive to the saddle; thus, the universal operator growth hypothesis must also include the saddle-dominated scrambling.

\section{Krylov complexity and saddle-dominated scrambling}\label{sec3}
In this section, we compute Krylov complexity in the same integrable system, namely the LMG model. The main objective is to demonstrate that in spite of being the system is integrable, K-complexity fails to identify the integrable nature of the system. This is due to the presence of saddle points in the phase space, which gives rise to the exponential growth of K-complexity at early times.

\subsection{Brief review of Krylov complexity}
\label{sec:review}
 
We start by introducing the operator growth and the K-complexity. Consider an operator $\mathcal{O}_{0}$ in a system governed by a time-independent Hamiltonian $H$. Under unitary time evolution, the time evolved operator $\mathcal{O}(t)$ is written via Heisenberg evolution as
\begin{equation}
    \mathcal{O}(t) = e^{i H t /\hbar}\,\mathcal{O}_{0}\,e^{-i H t / \hbar}\,.
\end{equation}
A standard way to compute the above time evolution is using the well-known Baker-Campbell-Hausdorff (BCH) formula, which involves increasingly complicated nested commutators. We use the following corollary to the BCH formula
\begin{equation}
    e^{X}Ye^{-X} = \sum_{n = 0}^{\infty} \frac{\mathcal{L}^{n}_{X}Y}{n!}\,,
\end{equation}
where we have used the Liouvillian super-operator defined as $\mathcal{L}_{X} Y = [X,Y]$. This gives us the known time evolution series for $\mathcal{O}(t)$
\begin{align}\label{bch}
    \mathcal{O}(t) = \mathcal{O}_{0} + \frac{i t}{\hbar} [H,\mathcal{O}] + \frac{(i t)^{2}}{2 !\,\hbar^{2}}[H, [H, \mathcal{O}]] + \frac{(i t)^{3}}{3 ! \, \hbar^{3}}[H, [H, [H, \mathcal{O}]]] + \cdots~~.
\end{align}
As time evolves, the complicated nested commutators are the indicators of the spreading of the initial operator. However, they become extremely difficult to compute. The prescription for calculating the K-complexity of such an operator growth stems from the need for an orthogonal basis created out of these nested operators
\begin{align}
    \mathcal{O}_{0} \equiv |\bar{\mathcal{O}}_{0}), \;\;\;\; \mathcal{L}^{1}_{H}\mathcal{O}_{0} \equiv |\bar{\mathcal{O}}_{1}), \;\;\;\; \mathcal{L}^{2}_{H}\mathcal{O}_{0} \equiv |\bar{\mathcal{O}}_{2}), \;\;\;\; \mathcal{L}^{3}_{H}\mathcal{O}_{0} \equiv |\bar{\mathcal{O}}_{3}), \;\;\;\; \cdots~~~.
\end{align}
Clearly, these operators do not form an orthogonal basis \textit{a priori}. Therefore, we perform an iterative Gram-Schmidt orthogonalization. For that, we need a definition of a norm of the said operators. A natural choice is the Wightman norm
\begin{align}
    (A|B) =  \langle e^{H \beta/2}A^{\dagger} e^{-H \beta /2} B \rangle_{\beta}\,,
\end{align}
where the $\langle \cdots \rangle _{\beta} = \mathrm{Tr}(e^{-\beta H} \cdots)/\mathrm{Tr}(e^{-\beta H})$ is the thermal expectation value at temperature $1/\beta$. However, in our discussion we will consider the infinite temperature inner product only\footnote{The finite temperature construction is itself an interesting problem. We leave this for future work.}, thus we set $\beta = 0$. The Gram-Schmidt orthogonalization procedure, better known in this case as the Lanczos algorithm, proceeds as follows (where we shall only consider the norm at $\beta$ = 0).\par 
 
\begin{itemize}
	\item Start with the definition $A_{0} \equiv \mathcal{O}_{0}$, which we assume to be normalized $(\mathcal{O}_{0}| \mathcal{O}_{0}) = 1 $. Normalization is defined via the Wightman norm (at $\beta = 0$), written as $(\mathcal{O}| \mathcal{O}') = \frac{1}{\mathcal{N} }\mathrm{Tr}(\mathcal{O}^{\dagger}\mathcal{O}' )$. 
	\item  Define $A_{1} = [H, \mathcal{O}_{0}]$, and normalize it with $b_{1} = \sqrt{(A_{1}|A_{1})}$. Define the normalized operator $\mathcal{O}_{1} = b_{1}^{-1}A_{1}$.
	\item From this, given $\mathcal{O}_{n-1}$ and $\mathcal{O}_{n-2}$, we can construct the following operators
	\begin{equation}
		A_{n} = [H, \mathcal{O}_{n-1}] - b_{n-1}\mathcal{O}_{n-2}\,.
	\end{equation}
	This can be normalized as $b_{n} = \sqrt{(A_{n}|A_{n})}$ and the $n^{\mathrm{th}}$ basis element is given by $\mathcal{O}_{n} = b_{n}^{-1}A_{n}$. 
	\item Stop the algorithm when $b_n$ hits zero.
\end{itemize}

The above algorithm provides a full orthonormal basis, known as the \textit{Krylov basis}. For a finite-dimensional system, the dimension of the Krylov basis (which we denote by $\mathcal{K}$) is finite. It obeys the bound \cite{Rabinovici:2020ryf}
\begin{align}
    1 \leq \mathcal{K} \leq D^2 - D + 1\,, \label{kdim}
\end{align}
where $D$ is the dimension of the Hilbert space under consideration. A chaotic system is expected to closely saturate the bound, whereas for an integrable system, $\mathcal{K}$ is much lower than the maximal bound.

Once the Krylov basis is prepared, the time-evolved operator $\mathcal{O}(t)$ can be expanded in this basis in the following way
\begin{equation} \label{Ot}
    |\mathcal{O}(t)) = \sum_{n = 0}^{\mathcal{K}-1} i^{n}\phi_{n}(t)|\mathcal{O}_{n})\,.
\end{equation}
The object of interest is the ``wavefunction'' $\phi_{n}(t)$. The Heisenberg time evolution equation of an operator implies $\partial_{t}\mathcal{O} = i[H,\mathcal{O}]$, and using the fact that the $\mathcal{O}_{n}$'s are orthonormal, we have the following ``recursion relation''
\begin{equation}
    \partial_{t}\phi_{n}(t) = b_{n}\phi_{n-1} - b_{n+1}\phi_{n+1}\,.
\end{equation}
This recursion can be visualized a particle hopping in a lattice where the hopping amplitudes are encoded in the Lanczos coefficients. The $\phi_{n}(t)$ can be evaluated also from \eqref{Ot} as
\begin{equation}
    \phi_{n}(t) = i^{-n}(\mathcal{O}_{n}|\mathcal{O}(t))\,,
\end{equation}
with $\phi_{-1} (t) = 0$ and $\phi_n (0) = \delta_{n 0}$. Further, $|\phi_n (t)|^2$ can be observed as the probabilities, and due to the unitary time evolution, we expect hence for any time $t$, we expect that probabilities are conserved, i.e.,
\begin{equation}
    \sum_{n=0}^{\mathcal{K}-1}|\phi_{n}(t)|^{2}=1\,. \label{prob}
\end{equation}
This is equivalent to the statement that $\partial_t \sum_n |\phi_n (t)|^2 = 0$.\par

We further define the average expectation of the particle hopping in the lattice site, which is nothing but the K-complexity
\begin{equation}
    K(t) = \sum_{n}n |\phi_{n}(t)|^{2}\,. \label{kkcom}
\end{equation}
It has been observed that for an integrable system, the Krylov complexity shows a power-law growth at early times, i.e.,
\begin{align}
    K(t) \sim (\alpha t)^{\frac{1}{1-\delta}}\,,
\end{align}
where $\alpha$ and $\delta$ are related to the asymptotic \emph{sub-linear} growth of Lanczos coefficients\footnote{As an example for $\delta = 1/2$, the probability amplitudes can be computed as $\phi_n (t) = \frac{(\alpha t)^n}{\sqrt{n !}} \exp(-\alpha^2 t^2/2)$ \cite{PhysRevE.104.034112}. A detailed list of $b_n$ and $K(t)$ for simple cases can be found in \cite{Fan:2022xaa}.}
\begin{align}
    b_n \sim \alpha n^{\delta}, ~~~~ 0 < \delta < 1\,. \label{sub}
\end{align}
For a quantum-chaotic system, K-complexity is supposed to demonstrate \textit{exponential} growth at early times, followed by linear growth and eventual saturation \cite{Barbon:2019wsy, Rabinovici:2020ryf}
\begin{align}
    K(t) \sim e^{2 \alpha t}\,.
\end{align}
This can be further traced back to the growth of Lanczos coefficients. For a chaotic system, the Lanczos coefficients show asymptotic-linear growth
\begin{align}
    b_n \sim \alpha n\,.
\end{align}
with the probability amplitudes being $\phi_n (t) = \tanh^n(\alpha t) \sech(\alpha t)$ \cite{Barbon:2019wsy}. A typical example is the SYK model which saturates the above bound with corresponding Lyapunov exponent given by $\lambda = 2 \alpha$ \cite{Parker:2018yvk}. This is supposed to hold for any operators, which has been hypothesised as the \emph{universal} growth of operators \cite{Parker:2018yvk}.

It is interesting to note the saturation value of K-complexity both for the integrable and non-integrable cases. At late times, for chaotic systems, it was argued in \cite{Rabinovici:2021qqt} that K-complexity should be close to $\mathcal{K}/2$, where $\mathcal{K}$ is the dimension of Krylov space. On the other hand, the integrable systems saturate the K-complexity much lower than $\mathcal{K}/2$. An interacting integrable system was studied in \cite{Rabinovici:2021qqt}; in spite of having an exponentially large Krylov dimension, K-complexity saturates a much lower value compared to a chaotic system.\par
The exponential behavior of K-complexity is hidden inside the auto-correlation function $C(t)$. In simple cases where the analytic form of the auto-correlation function is known, the Lanczos coefficients can be determined by a recursive method \cite{viswanath1994recursion, Parker:2018yvk}. However, for numerical calculations, this recursive methods suffers high instability. On the other hand, one can take the Fourier transform of the auto-correlation function and obtain the spectral function
\begin{align}
    \tilde{C}(\tilde{\omega}) = \int_{-\infty}^{\infty} \mathrm{d} t ~ e^{- i \tilde{\omega} t} C(t)\,. \label{spec}
\end{align}
With $b_n \sim \alpha n$, the spectral function decays exponentially as $\tilde{C}(\tilde{\omega}) \sim e^{-\pi|\tilde{\omega}|/2 \alpha}$. This can be obtained by observing $C(t) = \phi_0(t) = \sech(\alpha t)$, and plugging this to Eq.\eqref{spec} with large $|\tilde{\omega}|$ limit.. This in turn suggests that $C(t)$ with $t \in \mathbb{C}$ has pole at purely imaginary axis at location $t = \pm i \pi/2 \alpha$ (nearest to the origin) \cite{Parker:2018yvk}. We will come back to structure in later sections.

\subsection{K-complexity in LMG model}
In this section, we compute K-complexity and the associated growth of the Lanczos coefficients in the LMG model. The goal is to observe the behavior and implication of saddle-dominated scrambling on the Lanczos coefficients as well as on the K-complexity.

\begin{figure}
		\centering
		\begin{subfigure}[b]{0.48\textwidth}
		\centering
		\includegraphics[width=\textwidth]{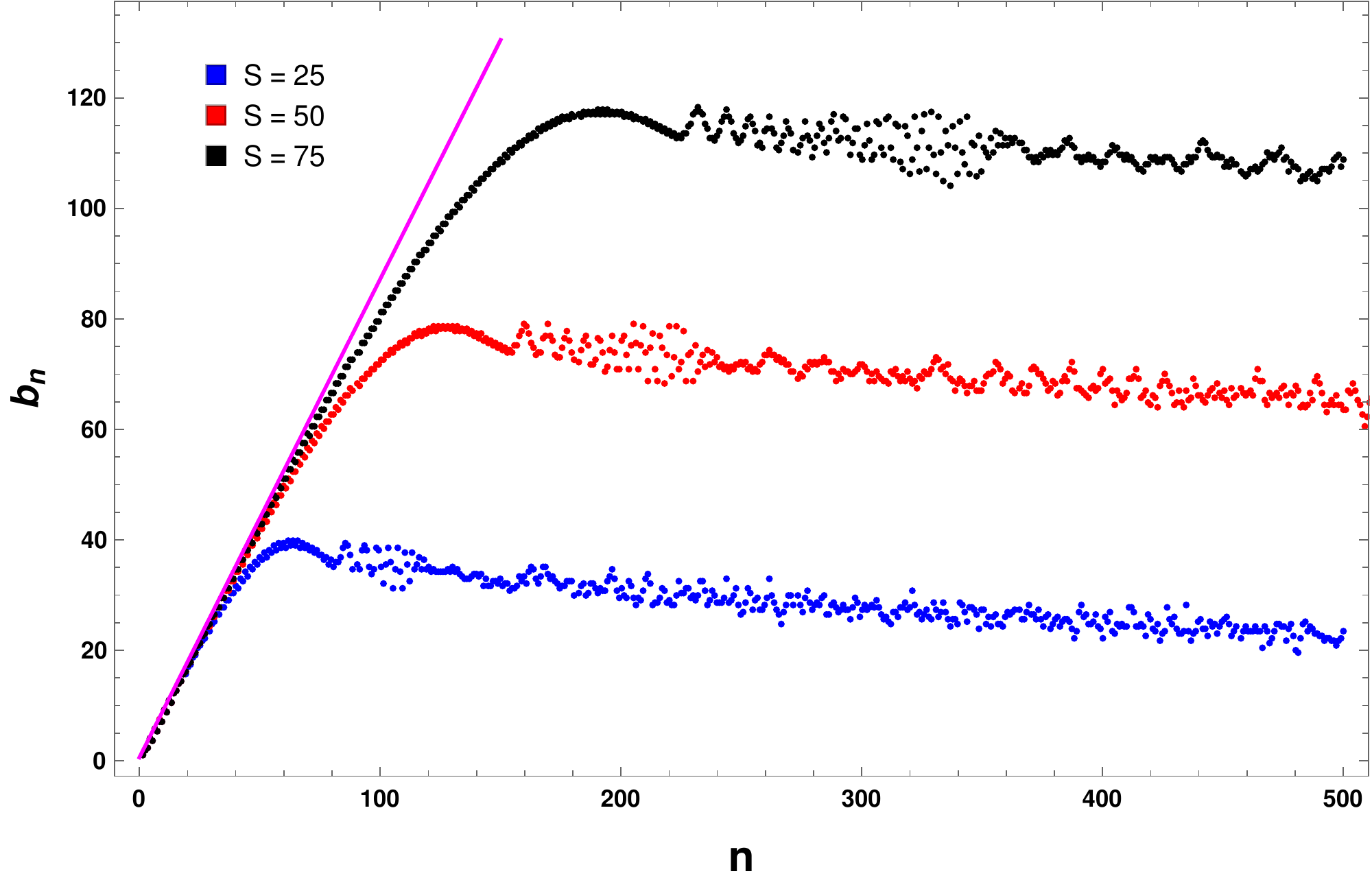}
			\caption{Growth of $b_{n}$ with $n$.}
			\label{fig:Kcomp}
		\end{subfigure}
		\hfill
		\begin{subfigure}[b]{0.48\textwidth}
		\centering
		\includegraphics[width=\textwidth]{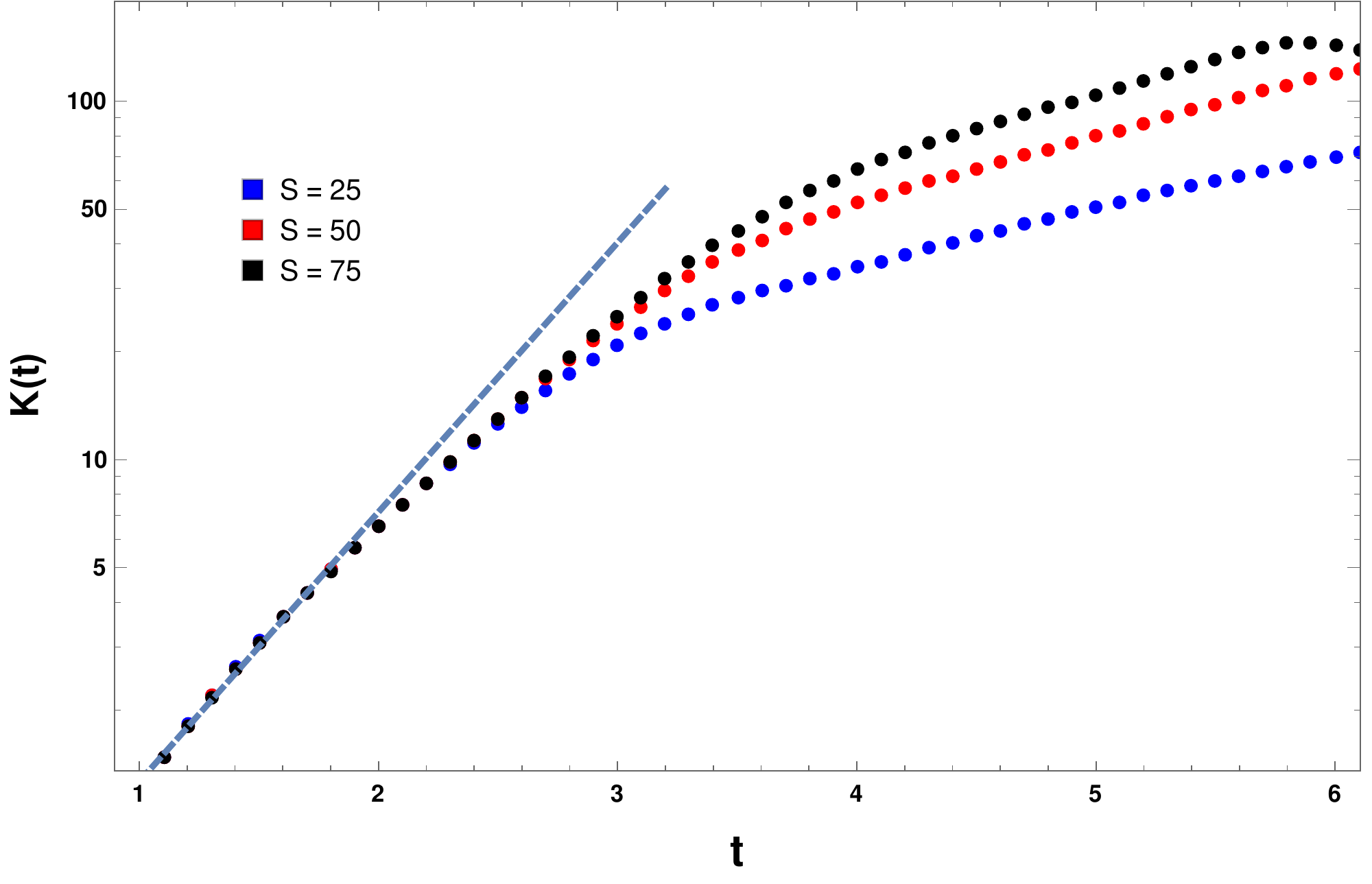}
			\caption{Growth of $K(t)$ with $t$ (log-plot).}
			\label{fig:Kent}
		\end{subfigure}
		%\hfill
		%\begin{subfigure}[b]{0.48\textwidth}
		%\centering
		%\includegraphics[width=\textwidth]{kent.pdf}
	%		\caption{Growth of $S_{K}(t)$ with $t$.}
		%	\label{fig:KentropyPlot}
		%\end{subfigure}
		\caption{(a) Behavior of Lanczos coefficients $b_{n}$ with $n$. At small $n$, the growth of $b_{n}$ is linear and indistinguishable for all spins. After some $n$ the higher spin saturates to a larger value of $b_{n}$ before eventually falling to zero. This is a consequence of the finite size of the system (b) Behavior of K-complexity (log-plot) with time. At early time growth is exponential, as expected from a chaotic system. After a time scale of the order of system size, the growth reduces to a power-law nature, before eventually saturating (at large times). While it is not evident from the figure, it can be seen that the K-complexities for higher spins saturate at higher values.} %(c) Behavior of K-entropy with time. The growth is linear at early times.}
		\label{figdiffop}
\end{figure}

As defined in the previous section, we compute the Lanczos coefficients and the Krylov complexity for the Hamiltonian \eqref{Hamquant}. We take the initial operator as $\mathcal{O}_0 = \hat{z}$ (similar to the OTOC case). Due to the presence of $\hbar_{\mathrm{eff}}$, one should, in principle, start from scratch and identify the steps where the $\hbar_{\mathrm{eff}}$ must be retained. However, we pick an alternate and simpler route by looking at the time evolution series for an operator $\mathcal{O}$, with the notation $\mathcal{O}(0) = \mathcal{O}_{0}$
\begin{align}
    \mathcal{O}(t) = \mathcal{O}_{0} + \frac{i  t}{\hbar_{\mathrm{eff}}}[H,\mathcal{O}_{0}] + \frac{ ( i t)^{2}}{\hbar^{2}_{\mathrm{eff}}}[H,[H,\mathcal{O}_0]] + \cdots,
\end{align}
where the ellipsis denotes the higher commutators. We simply absorb the $\hbar_{\mathrm{eff}}$ into Hamiltonian $H$. Therefore in terms of the rescaled Hamiltonian $\tilde{H} = H/\hbar_{\mathrm{eff}}$, the Lanczos algorithm follows through. Alternatively, it corresponds to the scaling of $b_{n}$'s according to
\begin{align}
    \tilde{b}_n = \frac{1}{\hbar_{\mathrm{eff}}}\, b_n = b_n \,S\,.
\end{align}
In the rest of the text, we use the familiar notation of $b_{n}$ to indicate $\tilde{b}_{n}$.
With this rectification,  we compute the Lanczos coefficients with various spins, namely for $S = 25, 50$, and $75$. 

\begin{figure}
		\centering
		\begin{subfigure}[b]{0.48\textwidth}
		\centering
		\includegraphics[width=\textwidth]{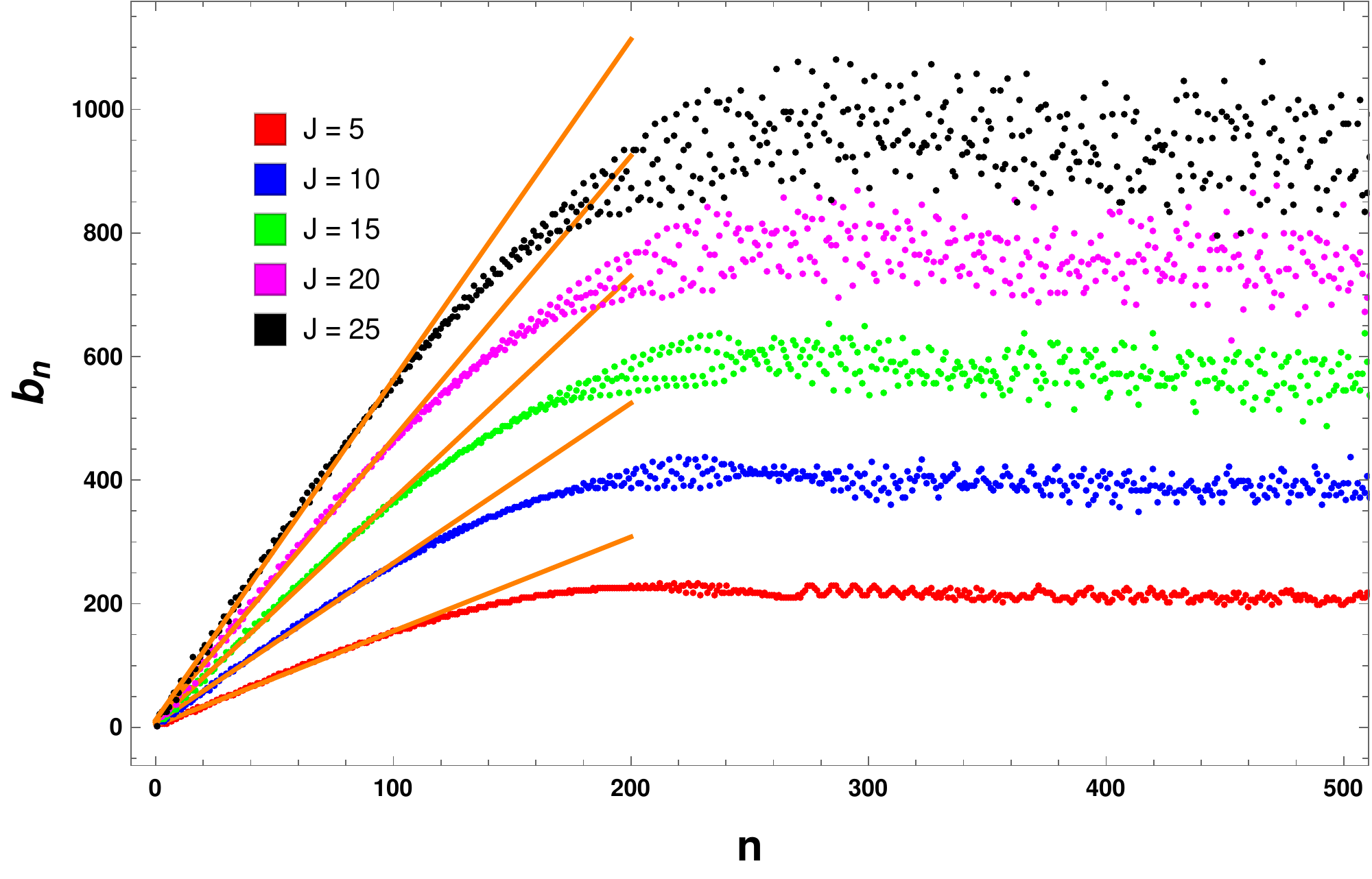}
			\caption{Growth of the $b_{n}$ with $n$ for various $J$ values in the $\text{LMG}$ model for spin $S = 75$.}
			\label{fig:jj}
		\end{subfigure}
		\hfill
		\begin{subfigure}[b]{0.48\textwidth}
		\centering
		\includegraphics[width=\textwidth]{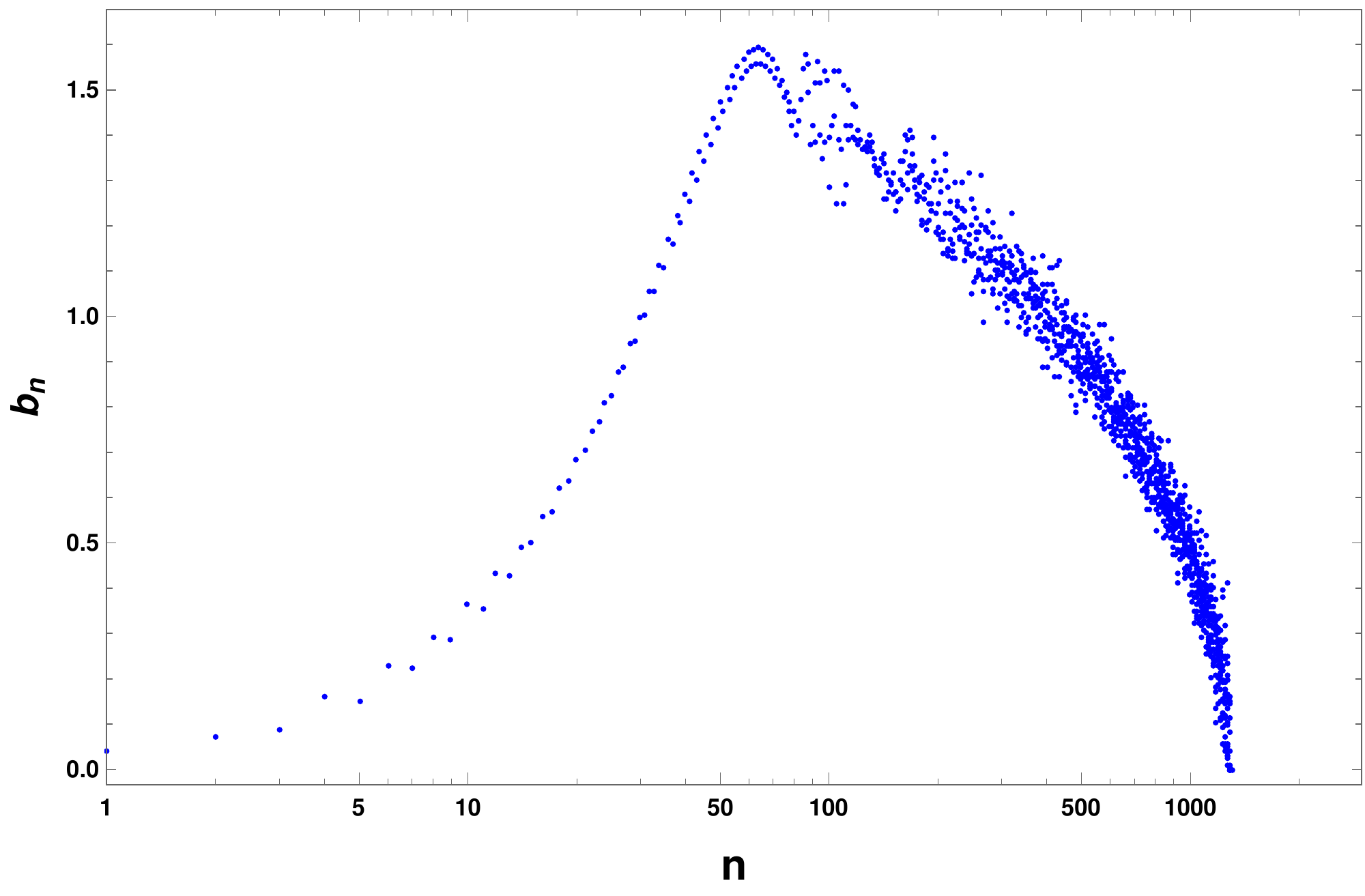}
			\caption{Growth of $b_n$ with $n$ in log-plot for $S = 25$ and $J = 2$.}
			\label{fig:jj2}
		\end{subfigure}
		\caption{(a) Behavior of the Lanczos coefficients for LMG model with spin  $S = 75$ and for $J = 5, 10, 15, 20, 25$. The straight lines have the equation $b_{n} = \alpha n + \beta$. The coefficient $\alpha$ follows the inequality $\alpha \geq \sqrt{2 J -1}/2$. (b) Growth of $b_n$ with $n$ in log-plot for $S=25$, and $J = 2$ till the edge of Krylov space.}
		\label{figJ}
\end{figure}

The results are shown in Fig.\,\ref{figdiffop}. As is evident from Fig.\,\ref{fig:Kcomp}, the Lanczos coefficients increase linearly for small $n$. This is the expected classical behavior due to the presence of an unstable saddle point in the phase-space of the classical LMG Hamiltonian. After some value of $n$, of the order of the spin $S$, finite-size effects kick in, and the Lanczos coefficients begin to saturate and eventually hit zero at the end of the Krylov space. In the region before $n \sim O(S)$, the Lanczos coefficients increase linearly with a slope $\alpha$ roughly equal to $\sqrt{3}/2$ (up to numerical precision). For large-$S$, the saturation takes place at a larger value. In principle, at $S \rightarrow \infty$, the saturation never occurs, and the Lanczos coefficients increase linearly with slope $\sqrt{3}/2$, which is expected at the classical limit.

 Equivalently, we can compute the K-complexity by first computing $\phi_n$'s and then using Eq.\eqref{kkcom}. %The $\phi_n$'s for first few $n$ are shown in Fig.\ref{fig:phin}. 
 Here we stress that for any time, the sum of the probabilities is conserved, and we have explicitly verified Eq.\eqref{prob} for all time. The behavior of K-complexity is shown in Fig.\,\ref{fig:Kent} (log-plot). The K-complexity grows exponentially with the exponent $2\alpha =  \sqrt{3}$, which is the same growth rate we observed for OTOC. This shows that an integrable system does not necessarily show the sub-linear growth of the Lanczos coefficients or the power-law growth of K-complexity. Even in an integrable system, K-complexity can grow exponentially, provided the analogous classical Hamiltonian possesses an unstable saddle point in the phase-space. This exponential growth is entirely due to the scrambling, and K-complexity is effective in capturing such phenomena. On the other hand, this also suggests that K-complexity is ignorant in distinguishing between the phenomena of saddle-dominated scrambling and chaos.
 
 Fig.\,\ref{fig:jj} shows the variation of Lanczos coefficients for different values of $J$ (with $S=75$) for the quantum analogue of the  Hamiltonian \eqref{LMGclassHam}. For each case, the Lanczos coefficients grow linearly with the slope determined by $J$. The growth rate $\alpha$ is lower-bounded as $\alpha = \sqrt{2J -1}/2$. Again, the equality is expected to saturate at the $S \rightarrow \infty$ limit. In Fig.\,\ref{fig:jj2}, we show the full variation of Lanczos coefficients (in log-plot) till the end of the Krylov space for $S=25$ and $J=2$. Here, the Krylov dimension is $\mathcal{K} \sim 1300$ which is lower than $D^2 - D + 1 = 2551$, supporting Eq.\eqref{kdim} \cite{Rabinovici:2020ryf, Rabinovici:2021qqt}. As discussed in the introduction, Lanczos ascent and descent are clearly visible. However, the plateau does not appear. This is due to the finite-dimensional size of the system as argued in previous computations, for example, in SYK model \cite{Rabinovici:2020ryf} and XXZ model \cite{Rabinovici:2021qqt}. However, there appears to a bump after the peak, which might be related to the integrability of the system.\footnote{We thank Pawel Caputa for drawing our attention to this point.} This requires further investigation.

\section{Microcanonical K-complexity}\label{sec4}

In the previous section, we have explicitly shown that due to the overwhelming contribution from the saddle, K-complexity exponentially increases at early times. This directs us to see whether we can use a finer probe near the saddle and detect contributions from that specific region. This allows us to examine the ``microcanonical K-complexity", which we compute near the saddle as well as the away from it. We see that we can accurately capture the behavior of microcanonical K-complexity both near and away from the saddle. Specifically, we will see that the exponential behavior holds not only near the saddle but also away from it. However, before delving into the details of the microcanonical case, we perform a classical analysis of K-complexity with fixed energies.

\subsection{Classical analysis of K-Complexity in LMG model}\label{sec41}
In this subsection, we present an analytical calculation of the K-complexity growth rate in the classical limit. The idea is to use the relation between $\alpha$ and the singularity locus of the auto-correlation function $C(t)$ in the complex plane $t \in \mathbb{C}$~\cite{Parker:2018yvk}. This can be in turn related to the singularities of the integral of motion, say $z(t)$, for $t\in  \mathbb{C}$, which can be analytically calculated. 

Consider a classical orbital of energy $E = H = x + J z^2$. Integrating the equation of motion $z'(t) = \{z, H\} = y$ by quadrature, we obtain 
\begin{align}
    t = \int \frac{\mathrm{d} z}{y_{E}(z)}\,,\label{eqt} 
\end{align}
where (recall that $x^{2} + y^{2} + z^{2} = 1$)
\begin{align}
    y_{E}(z) = \sqrt{ 1 - E^{2} + (2 J E - 1)z^{2} - J^{2} z^{4} } \,.
\end{align}
Eq. \eqref{eqt} should be viewed as an implicit way of defining $z_{E}(t)$. We now locate the singularity in $z_{E}(t)$ for complex value of $t$. Note that $z_{E}(t)$ is analytic for $t \in \mathbb{R}$. Imagine \eqref{eqt} as an equation of the form $t = F_{E}(z)$. Therefore, we would have $z = F^{-1}_{E}(t)$. The singularity in $z$ occurs when $F^{-1}_{E}(t)$ diverges. From that, it is not difficult to see that the imaginary part of the  singularity closest to the real plane has imaginary, $\sigma_*$, is given by
\begin{equation}
\sigma_* = \int_{z_{0}}^{\infty}\frac{\mathrm{d} z}{i y_{E}(z)} = 
    \frac{ \sqrt{2} \,\mathsf{K}\left(\frac{1-2 E J +\sqrt{4 J^2-4 E J+1}}{1-2 E J -\sqrt{4 J^2-4 E J+1}}\right)}{\sqrt{\sqrt{4 J^2-4 E J+1}+2 E J-1}} \,.  \label{eqt1.5}
\end{equation}
Here, $z_{0}$ is the zero of $y^{2}_{E}(z)$ with the largest real part, and $\mathsf{K}$ is a complete elliptic integral of the first kind. \par

Next, we relate $\sigma_*$ to the singularities of the auto-correlation function $C(t) = \left< z(0) z(t) \right>$ in the \textit{microcanonical} ensemble. For this, we replace the ensemble average by the time average
\begin{eqnarray}
C(t) = \frac1T \int z(s + t/2) z(s - t/2) \,\mathrm{d} s \,,
\end{eqnarray}
where the integral is over a period $T$. Now, if $z(s)$ is analytical in the strip $\{ s: \Im (s) \le \sigma_* \}$, $C(t)$ is analytical in the strip $\{ t:  \Im (t) \le \tau_* =  2 \sigma_* \}$, which has \textit{twice the width}. Now, it is known~\cite{Parker:2018yvk} that $\tau_*$ is related to the K-complexity growth rate by $\tau_* = \pi / (2\alpha)$. Combining with \eqref{eqt1.5}, we have finally
\begin{equation}
    \alpha(E) = \frac{\pi}{2 \tau_*} = \frac{\pi}{4 \sigma_*} =  \frac{\pi \sqrt{\sqrt{4 J^2-4 E J+1}+2 E J-1}}{ 4 \sqrt{2} \,\mathsf{K}\left(\frac{1-2 E J +\sqrt{4 J^2-4 E J+1}}{1-2 E J -\sqrt{4 J^2-4 E J+1}}\right)}\,. \label{eqt2}
\end{equation}
Here $\alpha(E)$ refers to the microcanonical ensemble with energy $E$.

We plotted \eqref{eqt2} as a function of $E$ for several values of $J$ in Fig.\,\ref{figmicrocanalpha}. We observe the following features: 
\begin{figure}[t]
		\centering
		\begin{subfigure}[b]{0.46\textwidth}
		\centering
		\includegraphics[width=\textwidth]{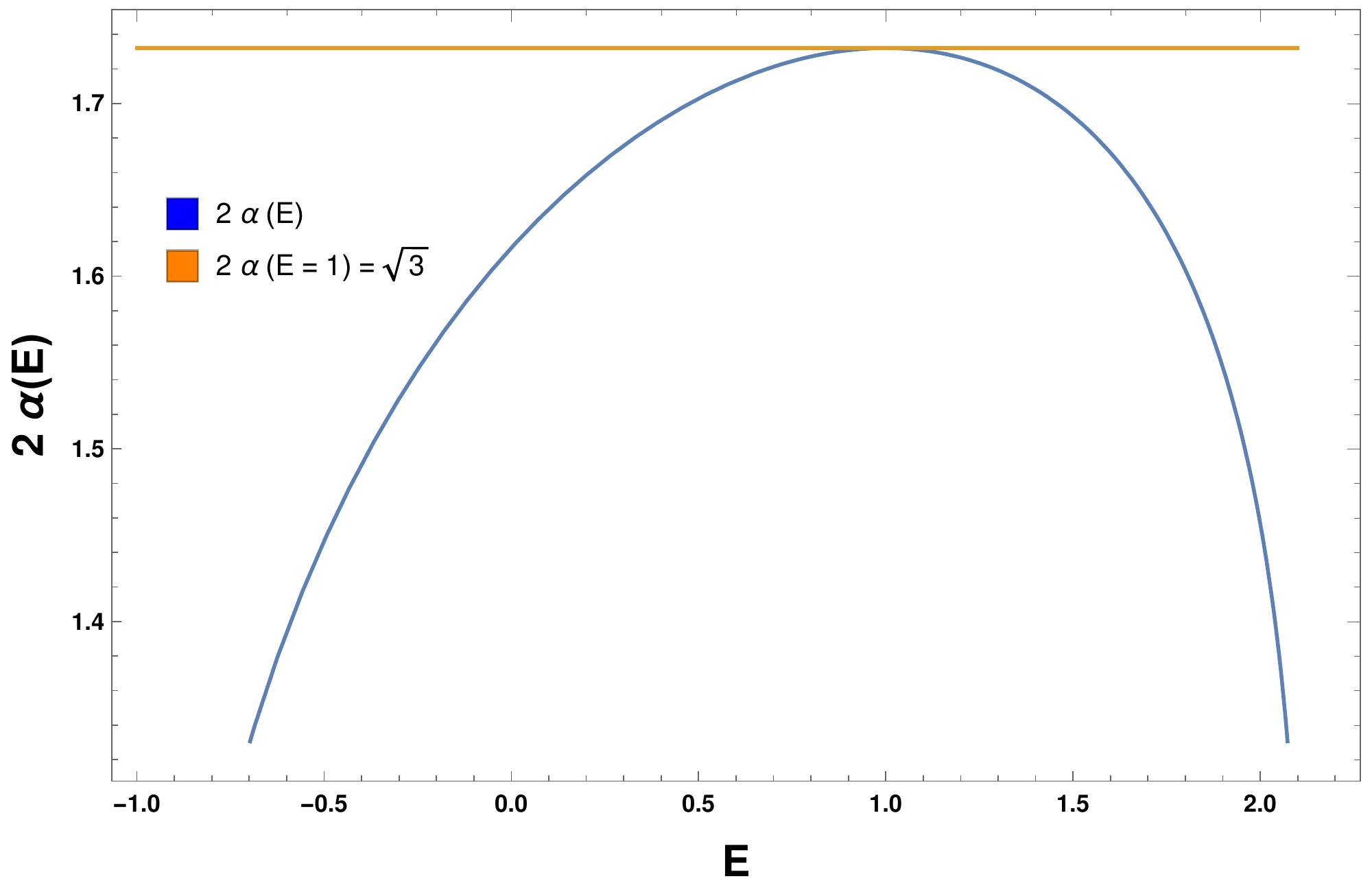}
			\caption{$2\alpha(E)$ for $J=2$.}
			\label{fig:j2}
		\end{subfigure}
		\hfill
		\begin{subfigure}[b]{0.46\textwidth}
		\centering
		\includegraphics[width=\textwidth]{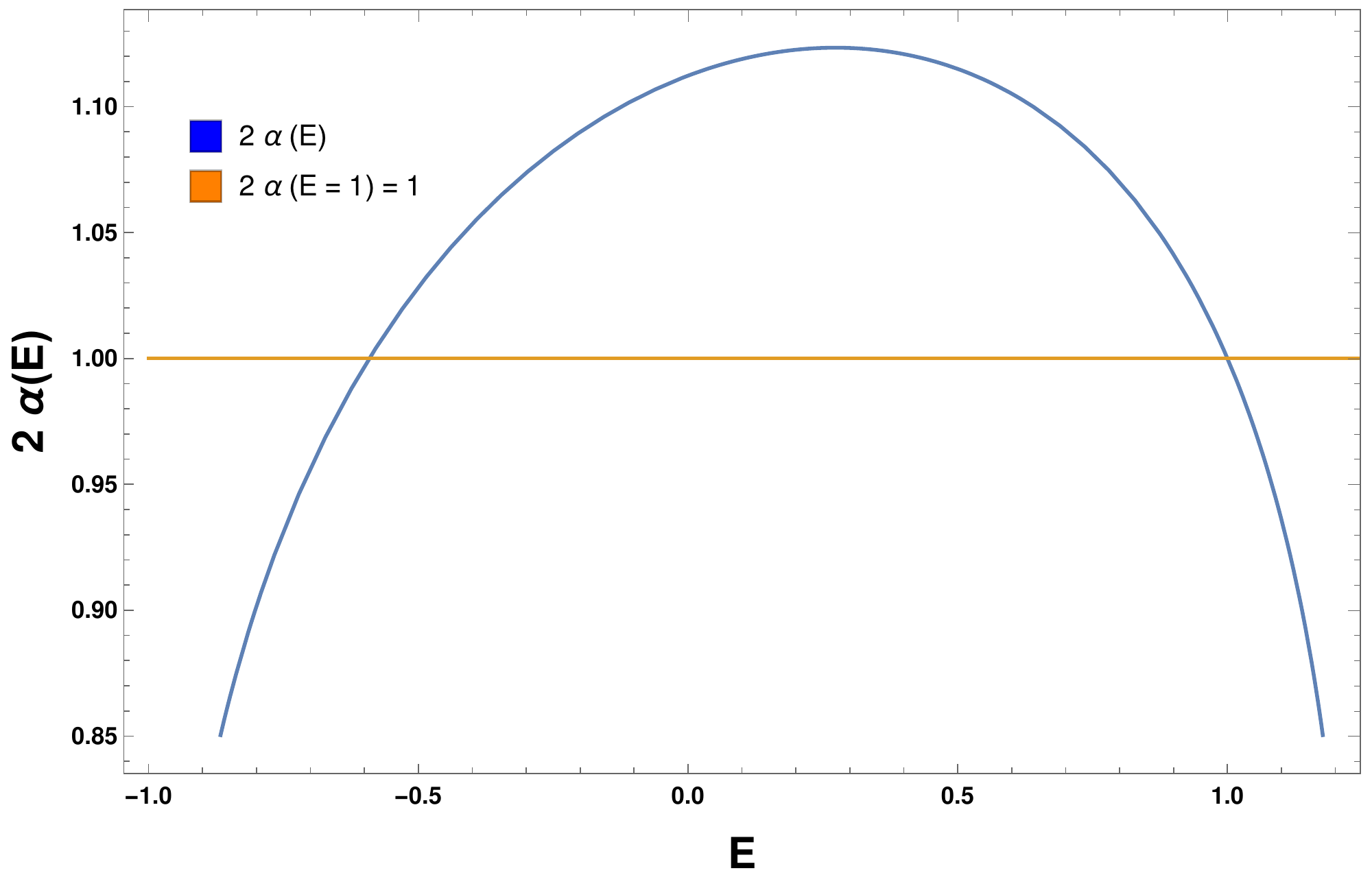}
			\caption{$2\alpha(E)$ for $J=1$.}
			\label{fig:j1}
		\end{subfigure}
		\hfill
		\begin{subfigure}[b]{0.46\textwidth}
		\centering
		\includegraphics[width=\textwidth]{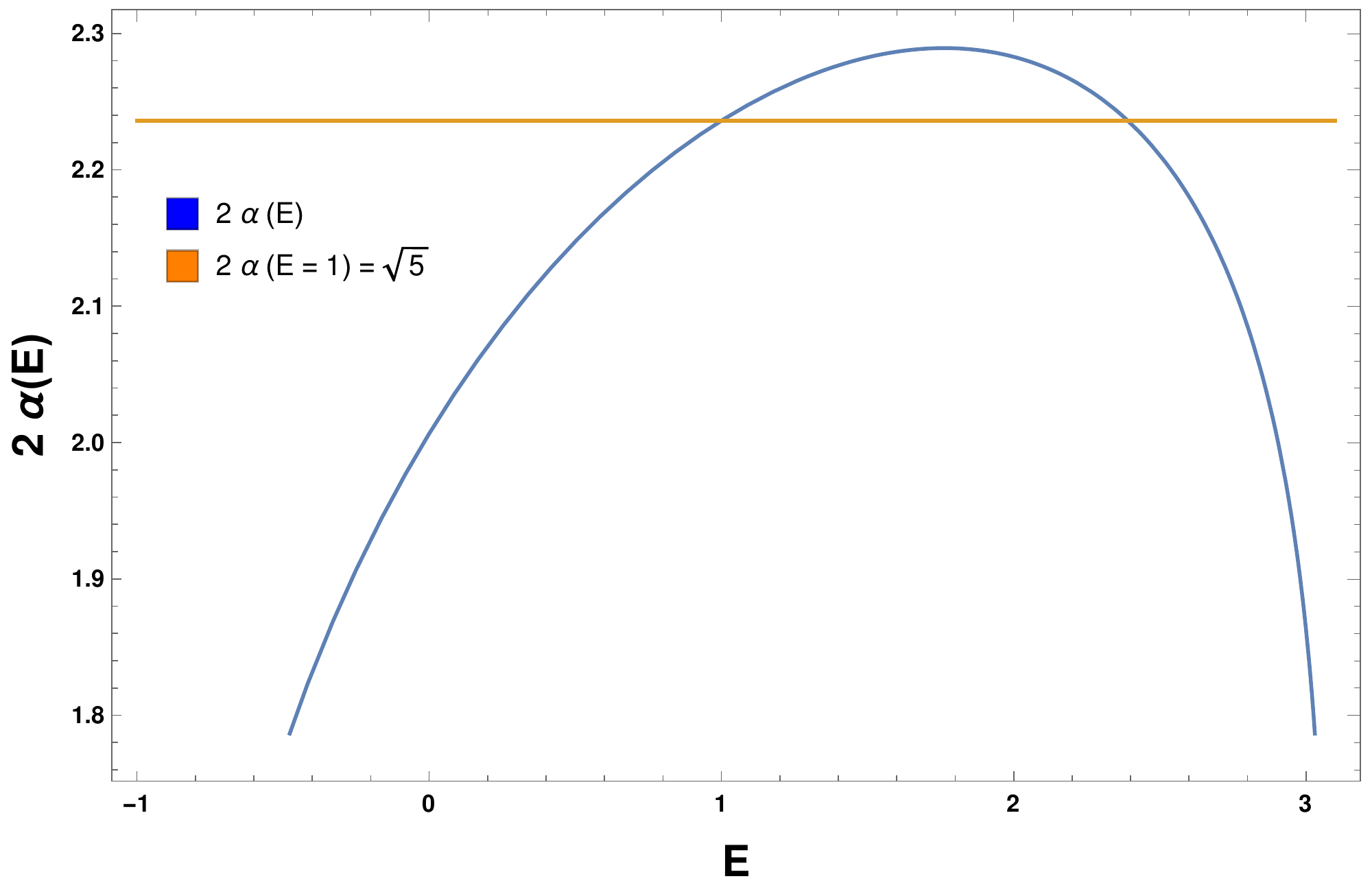}
			\caption{$2\alpha(E)$ for $J=3$.}
			\label{fig:j3}
		\end{subfigure}
		\hfill
		\begin{subfigure}[b]{0.46\textwidth}
		\centering
		\includegraphics[width=\textwidth]{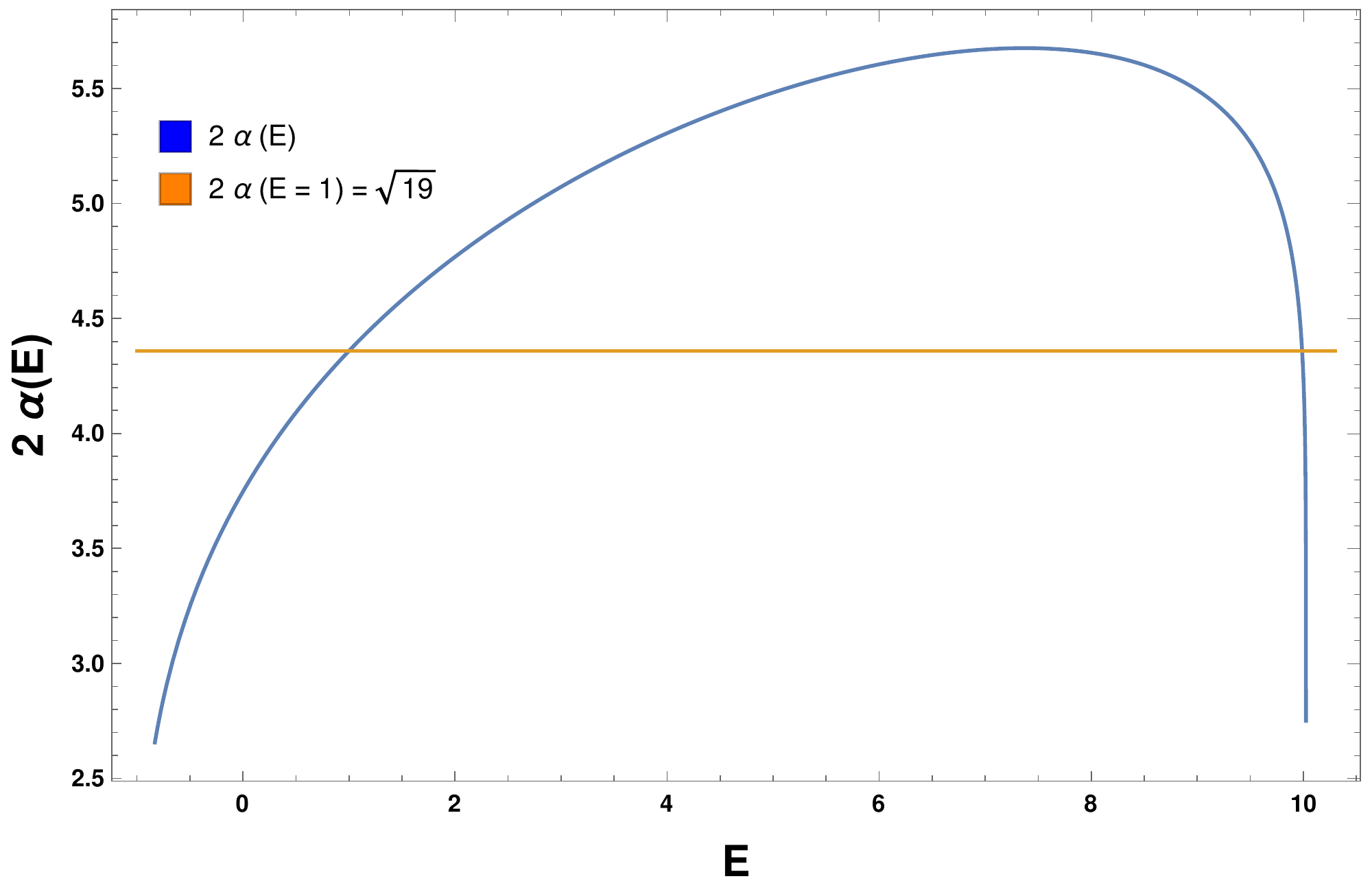}
			\caption{$2\alpha(E)$ for $J=10$.}
			\label{fig:j10}
		\end{subfigure}
		\caption{(a) Behavior of $2\alpha$ with respect to the energy $E$ plotted for $J = 2$ (blue). The $2\alpha (E = E_{\text{saddle}}) = \sqrt{2 J -1} = \sqrt{3}$ is given in orange. As is evident, the inequality $\sup_{E} 2\alpha(E) \geq \sqrt{2 J - 1}$ is saturated in this case. (b) Behavior of $2\alpha$ with respect to the energy $E$ plotted for $J = 1$ (blue). The $2\alpha (E = E_{\text{saddle}}) = \sqrt{2 J -1} = 1$ is given in orange. The inequality holds, and is not saturated. (c) and (d) Behavior of $2\alpha$ with respect to the energy $E$ plotted for $J = 3$ and $10$ respectively (blue). The $2\alpha (E = E_{\text{saddle}}) = \sqrt{2 J -1} = \sqrt{5}$ and $\sqrt{19}$ respectively, and are given in orange. The inequality holds, and is not saturated. In fact, we can see that the difference between the two values increases with $J$. Note that this happens for $J > 2$. For $1/2 <J < 2$, the difference decreases as $J$ increases until it goes to $0$ at $J = 2$.}
		\label{figmicrocanalpha}
\end{figure}

\begin{enumerate}
    \item When $E =  E_{\text{saddle}} = 1$ is equal to the saddle-point energy
    \begin{equation}
    2 \alpha(E = E_{\text{saddle}})=
      \sqrt{2 J - 1} =  \omega_{\text{saddle}}  \,, \label{alsadd}
    \end{equation}
    where $\omega$ is the leading eigenvalue of the linearized dynamical matrix at the saddle. The microcanonical $2 \alpha$ of the saddle energy level saturates the bound.
    \item In general, $2 \alpha(E)$ is non-vanishing for all energy (present in the system), and the maximum is away from $E = 1$, and is generally larger than $ \omega_{\text{saddle}}$, i.e.,
    \begin{equation}
        2 \,\mathrm{sup}_{E} \alpha(E) \ge \omega_{\text{saddle}}\,.
    \end{equation}
    Coincidentally, at $J = 2$ (the value we have chosen) this inequality is saturated, as demonstrated in Fig.\,\ref{fig:j2}. For $J > 2$ ($J < 2$), the minima satisfies $E > 1$ ($E < 1$), respectively shown in Fig.\,\ref{fig:j1} - Fig.\,\ref{fig:j10}. 
\end{enumerate}

\subsection{Microcanonical behavior near the saddle}
In this subsection, we explore the behavior of K-complexity for certain energy eigenvalues of the Hamiltonian. This amounts to evaluating the K-complexity and the corresponding Lanczos coefficients at some fixed energy. We call this microcanonical K-complexity (in analogy to the microcanonical OTOC). A similar refined version of the K-complexity for fixed average energy was previously introduced in \cite{Kar:2021nbm}. The average energy operator $\mathcal{E}$ is defined as $\mathcal{E} \ket{O} = 1/2\ket{\{H,O\}}$, and it commutes with the Liouvillian operator $[\mathcal{L}, \mathcal{E}] = 0$. Hence, the operator $\mathcal{E}$ acts as a conserved quantity for $\mathcal{L}$, and the whole Lanczos algorithm can be implemented for a particular energy sector. The associated Lanczos coefficients and K-complexity can be well defined in such a sector, and that is what we will exactly compute.

\begin{figure}[t]
		\centering
		\begin{subfigure}[b]{0.46\textwidth}
		\centering
		\includegraphics[width=\textwidth]{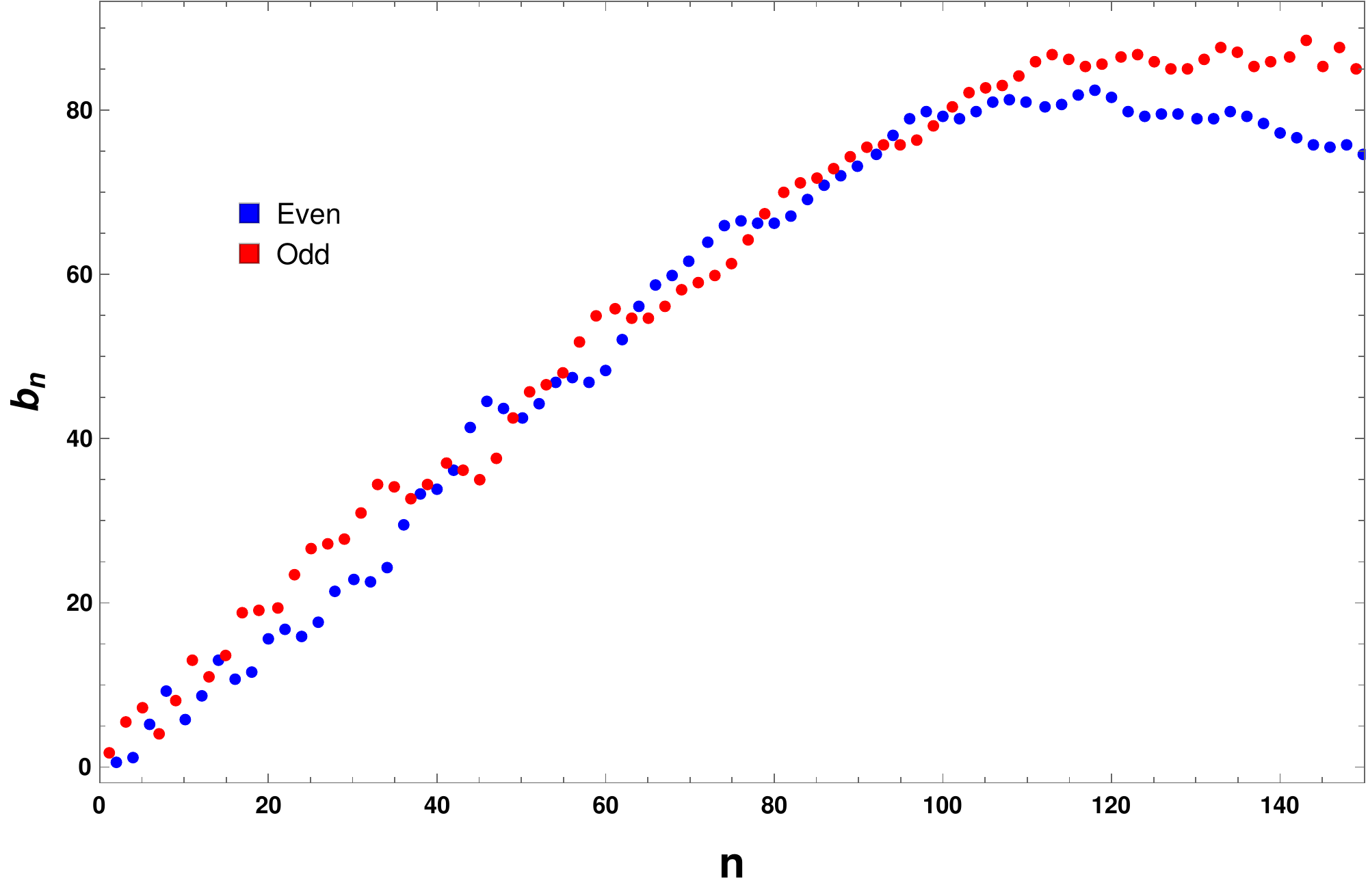}
			\caption{Lanczos coefficients for $E \sim 0$.}
			\label{fig:e0}
		\end{subfigure}
		\hfill
		\begin{subfigure}[b]{0.46\textwidth}
		\centering
		\includegraphics[width=\textwidth]{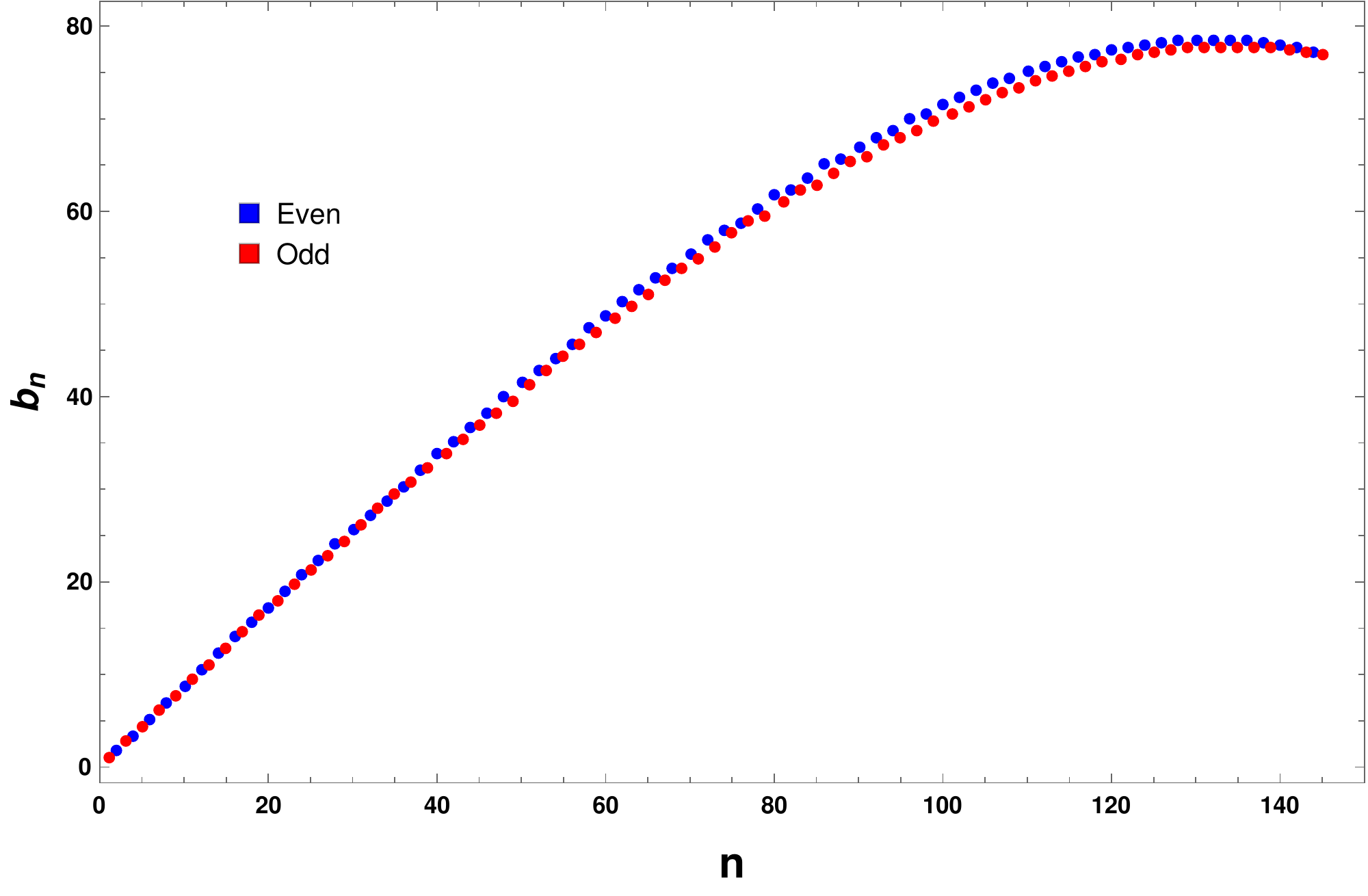}
			\caption{Lanczos coefficients for $E \sim 1$.}
			\label{fig:e1}
		\end{subfigure}
		\hfill
		%\hfill
		%\begin{subfigure}[b]{0.46\textwidth}
		%\centering
		%\includegraphics[width=\textwidth]{microkcom.pdf}
		%	\caption{K-complexity for different $E$.}
		%	\label{fig:mick}
		%\end{subfigure}
		\caption{Growth of Lanczos coefficients and microcanonical K-complexities for different energies. (a) shows the behavior away from the saddle where the odd and even $b_n$'s are clearly distinguishable. At small $n$ one can observe that the growth is roughly linear, which is indicative of an exponentially increasing K-complexity. Therefore, the microcanonical K-complexity is also sensitive to the presence of the unstable saddle point, even away from the saddle itself. (b) shows the growth of $b_n$ near the saddle. The odd and even coefficients are almost indistinguishable.
		}
		\label{figmicrocansad}
\end{figure}

Evaluating the microcanonical K-complexity is useful. The main advantage is that it can be used as a finer probe and capture the exact behavior near the saddle point rather than averaging out of the whole phase space. For this purpose, we replace the definition of the inner product with the sum over the desired eigenvectors according to
\begin{align}
    (A|B)= \frac{1}{N}\sum_{\{E-\Delta E, E + \Delta E\}}\bra{n} A^{\dagger}B \ket{n},
\end{align}
where the summation is performed with the eigenvalues corresponding to the energy $E$ and $N$ is the total number of eigenstates that are averaged over. For the averaging process, we take the energies within a window of $\Delta E$. This is in a similar spirit to the computation performed for the case of microcanonical OTOC \cite{Hashimoto:2017oit, Hashimoto:2020xfr}. With this similar motivation, here we study the behavior of microcanonical K-complexity for eigenvalues corresponding to the saddle-point energy $E \sim 1$, and one other case which is away from the saddle, namely, $E \sim 0$. For this, we choose the eigenvectors corresponding to the energy eigenvalues within a window of $\Delta E = \pm 0.1$. The Fig.\,\ref{fig:e0} and Fig.\,\ref{fig:e1} shows the behavior of the Lanczos coefficients for two different energies, one is at the saddle and other one is away from it. It is interesting to note that for the classical saddle ($E\sim1$), the odd and even Lanczos coefficients have the same behavior as can be seen from the Fig.\,\ref{fig:e1}. On the other hand, for the energy eigenvalues away from the saddle, we observe that the odd and even Lanczos coefficients can be distinguished (see Fig.\,\ref{fig:e0}). A heuristic explanation for the same is provided in the next subsection.\par

From the Lanczos coefficients of two different energies, it is straightforward to compute the K-complexities. We immediately find that the dominant behavior to the total K-complexity (calculated in previous sections) comes from the energy eigenvectors, which are near to the saddle $E \sim 1$. Corresponding to this energy, the Lanczos coefficients grows linearly at small $n$ (and so K-complexity would be exponential at early times), which is again distinguishable from the energies away from the saddle. \par
It is worth emphasizing that even away from the unstable saddle (which is characterized by the energy $E \sim 1$), the growth of the Lanczos coefficients is linear for small values of $n$, and correspondingly the K-complexity shows exponential growth at early times. Therefore, the microcanonical K-complexity detects the effect of the unstable saddle even away from the saddle itself. We can analytically observe the same by noting that the auto-correlation function has a purely complex pole $\alpha (E)$ for all values of $E$ lying between the maximum and minimum possible values. The value of $\alpha(E)$ satisfies the bound $2 \sup_{E}\alpha(E) \geq \omega_{\text{saddle}}$.

\subsection{Oscillations in Lanczos coefficients}

Here, we provide a plausible explanation of the different behavior of odd and even Lanczos coefficients that we have come across in previous discussions. As we will see, it can be traced back to the late-time behavior of the auto-correlation function.\par
Consider the Lanczos coefficients with oscillations \cite{PhysRevLett.124.206803, PhysRevB.102.195419}
\begin{equation}
    b_n = f(n) + (-1)^n g(n)\,,
\end{equation}
where $f(n)$ and $g(n)$ are rather slowly varying functions. For even and odd $n$, the Lanczos coefficients will oscillate between $f(n) \pm g(n)$. Let us also assume that $g(n) \ll f(n)$ when $n$ is large, and $g(n)$ decays with $n$. We consider the Schr\"odinger equation on the Krylov chain:
\begin{equation}
    \dot\varphi_n = -b_{n+1} \varphi_{n+1}+b_n \varphi_{n-1} \,.
\end{equation}
We plug in an ansatz 
\begin{equation}
    \varphi_n = \phi(n) + (-1)^n \psi(n)\,,
\end{equation}
where $\phi$ and $\psi$ are slowly varying in $n$ (this is an expansion around $0$ and $\pi$). As a result we have 
\begin{equation}
   \frac{\mathrm{d}}{\mathrm{d}t} \begin{pmatrix} \phi \\ \psi \end{pmatrix}  = 2  \begin{pmatrix}  - f \partial_n & -g \\ g & f \partial_n  \end{pmatrix}\begin{pmatrix} \phi \\ \psi \end{pmatrix}.
\end{equation}
In terms of a rescaled variable 
\begin{equation}
    u = \int \frac{\mathrm{d} n}{2 f(n)}\,,
\end{equation}
we can express the above equation as
\begin{equation}
 ( \partial_t + \partial_u ) \phi =2  g \psi \,,\, ~~~
 ( \partial_t - \partial_u ) \psi = - 2 g \phi  \,.
\end{equation}

\begin{figure}[t]
		\centering
		\includegraphics[width=0.67\textwidth]{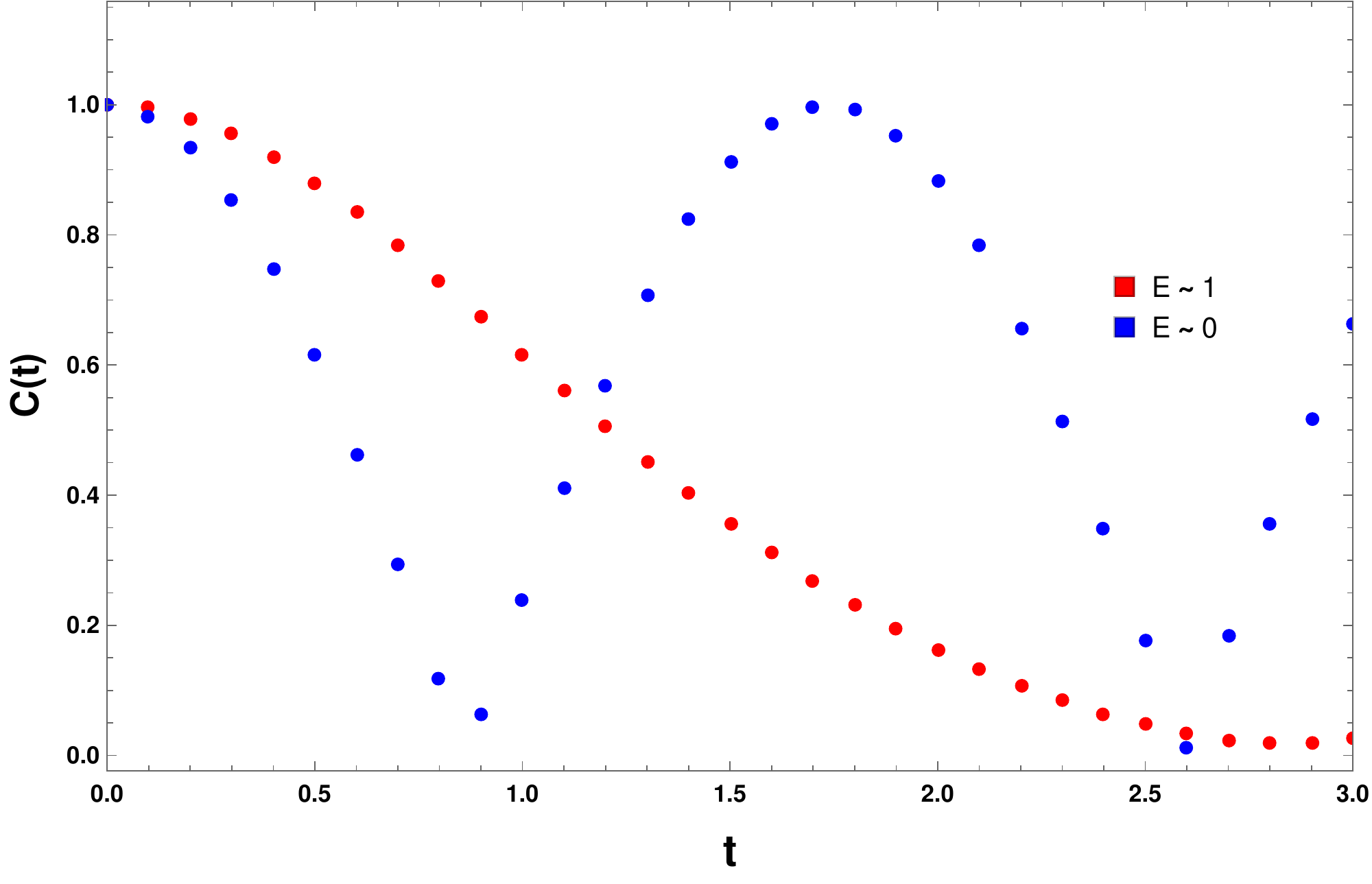}
		\caption{The behavior of the microcanonical auto-correlation function $C(t)$ with $t$ for different values of energy $E$. The function $C(t)$ decays rapidly for $E \sim 1$ and shows oscillatory for $E \sim 0$. This behavior is consistent with the fact that the even and odd Lanczos coefficients behave differently away from the saddle. In contrast, their growth is very similar at the saddle and nearly indistinguishable.}
		\label{fig:autocorr}
\end{figure}

We now treat $g$ perturbatively (since we assumed that $g$ is small for $u$ large), to the first order, focusing on the auto-correlation $\varphi_0(t) = \psi(0) + \phi(0)$. Recall the initial condition $\varphi_n(t=0) = \delta_{n,0}$. At zeroth order, $\phi$ describes a wavefront travelling along the line $u = t + C$ (towards $n\to\infty$), and $\psi$ describes a wavefront travelling along the line $u = -t + C$ (from $n \to \infty$). At $O(g)$, we can have one deflection $\phi \to \psi$ (or vice versa).  So the leading contribution to $(u = 0, t)$ has a trajectory $(0,0) \to (u= t/2,t/2) \to (0, t)$, hence
\begin{align}
    \varphi_0(t) \sim  g(n(u = t/2))   + O(g^2) \,.
\end{align} 
As an example, consider $f(n) = \alpha n$, and thus $u = \frac{1}{2\alpha} \ln n$. This implies $n = e^{2 \alpha u}$. If we have $g(n) \sim  (\ln n)^{-a}$ with $a>0$, then the auto-correlation function will have the form
\begin{align}
     C(t) = \varphi_0(t) \sim t^{-a} \,.
\end{align}
Therefore, in a system with linear Lanczos coefficient growth, a power-law decay of the auto-correlation function can come from a logarithmically decaying oscillation of the Lanczos coefficients on top of the growth. In particular, setting $a = 0$, we see that if there is a constant oscillation $g(n) \sim \text{constant}$, $C(t)$ does not decay to zero.

Coming back to our calculation, we compute the microcanonical version of the auto-correlation function for two different energies, $E \sim 1$ (saddle) and $E \sim 0$ (away from saddle). The result is shown in Fig.\,\ref{fig:autocorr}. We see that at the saddle-point, the auto-correlation decays to zero, while away from the saddle, it does not decay to zero (the precise nature is not very important for us). This explicitly confirms our conclusion, odd and even Lanczos coefficients are nearly indistinguishable at the saddle, while they show some small oscillations and can be distinguishable away from the saddle.

\section{Conclusions and outlook}\label{conclusion}
In this paper, we have performed a detailed analysis of K-complexity and the associated Lanczos growth in an integrable system exhibiting saddle-dominated scrambling. We find that the expectation of the sublinear growth of Lanczos coefficients (and the power-law growth of K-complexity), according to the universal operator growth hypothesis \cite{Parker:2018yvk}, does not meet. However, it does not imply the failure of the hypothesis, which states that the chaotic systems exhibit the fastest (linear) growth of the Lanczos coefficients. However, the inverse of the statement, i.e. any system demonstrating linear Lanczos growth is chaotic; need not be true. There might be, in principle, integrable systems which show similar growth. For example, the LMG model which we have studied in this paper belongs to this class. We have indeed found that Lanczos coefficients grow linearly, which gives rise to the early-time exponential growth of the K-complexity. This is reminiscent of their expected behavior for chaotic systems. As we have discussed in detail, this is entirely due to the presence of the unstable saddle point in the classical phase space. The saddle dominates not only a particular region but also the overall phase space. This can be seen via the microcanonical version of the K-complexity. The growth is still exponential away from the saddle, albeit slower than the saddle.
The overall qualitative nature of the K-complexity suggests, at least from the example we studied, that the K-complexity is very sensitive to the saddle point. Whenever such points exist, K-complexity fails to distinguish a saddle-dominated scrambling and generic chaos. It is still an open question about the generic nature of such behavior, and it will be interesting to see whether it can be exhibited by other $q$-complexities \cite{Parker:2018yvk}.

However, there is a positive sign of hope. In a purely chaotic system (in the absence of any saddle), it is reasonable to argue that the microcanonical K-complexity would give the same exponent irrespective of the subregion in the classical phase space we choose. However, this is not the case in the presence of an unstable saddle. The exponential growth of microcanonical K-complexity away from the saddle is much slower than the saddle itself. Hence, in this spirit, the microcanonical K-complexity is actually detecting the saddle. However, to detect this, we require high precision. Therefore, from a global perspective, it might be possible to construct some variant of K-complexity that utilizes this feature to distinguish between saddle-dominated scrambling and chaos.

We also observe some interesting features of the odd and even coefficients in the microcanonical version. The coefficients overlap at the saddle but show some oscillatory behavior away from it. We have provided a qualitative analysis of this observation from the nature of the auto-correlation function. The auto-correlation decays exponentially at the saddle and can be associated with the smooth growth of Lanczos coefficients at the saddle. We believe this is a generic feature of any model (at least for finite-dimensional systems) which can be confirmed by studying other many-body systems.

The saddle-dominated scrambling is not limited to the LMG model; instead, one can consider a variety of models of such behavior, such as the quantum Dicke model \cite{PhysRev.93.99} where substantial works have been dedicated for studying scrambling and OTOC \cite{Xu:2019lhc, Chavez-Carlos:2018ijc, Lewis-Swan:2018sdr, Wang:2018tmi}. Another simple example will be to consider the inverted harmonic oscillator. The system is not chaotic but shows instability \cite{Hashimoto:2020xfr, Bhattacharyya:2020art}. Similar conclusions are expected to hold for K-complexity and its microcanonical versions, where the dominating behavior is supposed to be controlled by the saddle. This will confirm whether such behavior is generic or not limited to a particular class of models. Furthermore, we have considered large-$S$ expansion with the classical limit being at $S \rightarrow \infty$. It is not clear that such models always possess a holographic dual. In such cases, it is imperative to consider the large-$N$ systems. It will be interesting to see whether saddle-dominated scrambling exists in such systems and, if so, then the role of K-complexity in those cases.

\section*{Acknowledgements}

We thank B. Ananthanarayan, Pawel Caputa, Shouvik Datta and Chethan Krishnan for comments on the draft. XC would like to thank Daniel Parker, Thomas Scaffidi and Tianrui Xu for discussions and collaborations in related topics. PN thanks Aranya Bhattacharya, Arpan Bhattacharyya, Pawel Caputa, Pingal Pratyush Nath and Aninda Sinha for useful discussions and ongoing collaborations. BB is supported by the Ministry of Human Resource Development (MHRD), Government of India through the Prime Ministers’ Research Fellowship. PN is supported by the University Grants Commission (UGC), Government of India.

\appendix
\section{Appendix: Feingold-Peres (FP) model}\label{appendFP}
In this appendix, we investigate the classical Feingold-Peres (FP) model of coupled tops \cite{FEINGOLD1983433, PhysRevA.30.504, PhysRevA.30.509}. This model shows the saddle-dominated scrambling despite being classically chaotic~\cite{Xu:2019lhc}. Meanwhile, the K-complexity growth rate is in general strictly greater than the saddle-point contribution~\cite{Parker:2018yvk}, and the value of $\alpha$ remains to be understood analytically.

\subsection{Classical saddles and Lanczos coefficients}\label{app1}
Let us first identify the saddle points. The Hamiltonian of the FP model is given by 
\begin{align}
    H = (1+c)(x_{1} + x_{2}) + 4 (1 - c)z_{1}z_{2}\,,
\end{align}
where $(x_{i},y_{i},z_{i})$ are two independent classical $\text{SU}(2)$ spins satisfying $x_{i}^{2} + y_{i}^{2} + z_{i}^{2} = 1$ and $\{x_{i},y_{i}\} = z_{i}$, with $i,j = 1,2$. The parameter $c \in [-1,1]$. The time evolution equations for this system (from Eq.\,\eqref{eqnom}) can be written as
\begin{align}
    \frac{\mathrm{d} x_{1,2}}{\mathrm{d} t} &= - 4 (1 - c)y_{1, 2}z_{2, 1}\,, \\
    \frac{\mathrm{d} y_{1, 2}}{\mathrm{d} t} &= - (1+c)z_{1,2} + 4(1-c)x_{1, 2}z_{2, 1}\,,\\
    \frac{\mathrm{d} y_{1, 2}}{\mathrm{d} t} &= (1 + c)y_{1, 2}\,.
\end{align}
At the saddles, we require the derivatives to vanish. These give us the following conditions
\begin{align}
    y_{1,2} &= 0\,,\\
    -(1+c)z_{1,2} + 4(1-c)x_{1,2}z_{2,1} &= 0\,. \label{fpsad}
\end{align}
A direct consequence of these relations is that $x_{1}x_{2} = \big(\frac{1 + c}{4(1-c)}\big)^{2} \leq 1$. This implies that saddle point exists only for $c\leq 3/5$. Hence, we restrict our discussion to $c \in [-1, 3/5]$.

\begin{figure}
		\centering
		\begin{subfigure}[b]{0.46\textwidth}
		\centering
		\includegraphics[width=\textwidth]{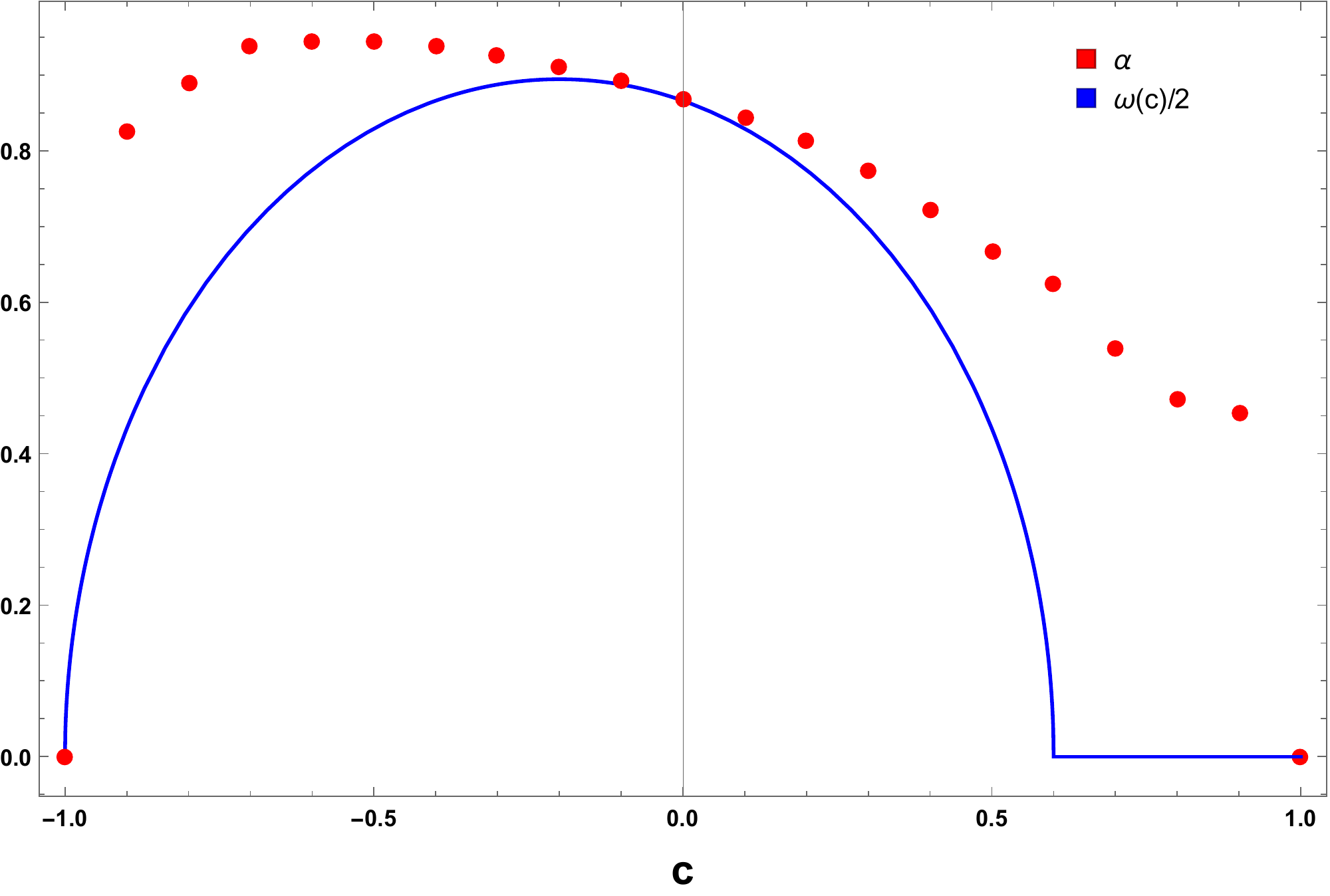}
			\caption{Variation of $\omega (c)/2$ and $\alpha$ with $c$.}
			\label{fig:omega}
		\end{subfigure}
		\hfill
		\begin{subfigure}[b]{0.48\textwidth}
		\centering
		\includegraphics[width=\textwidth]{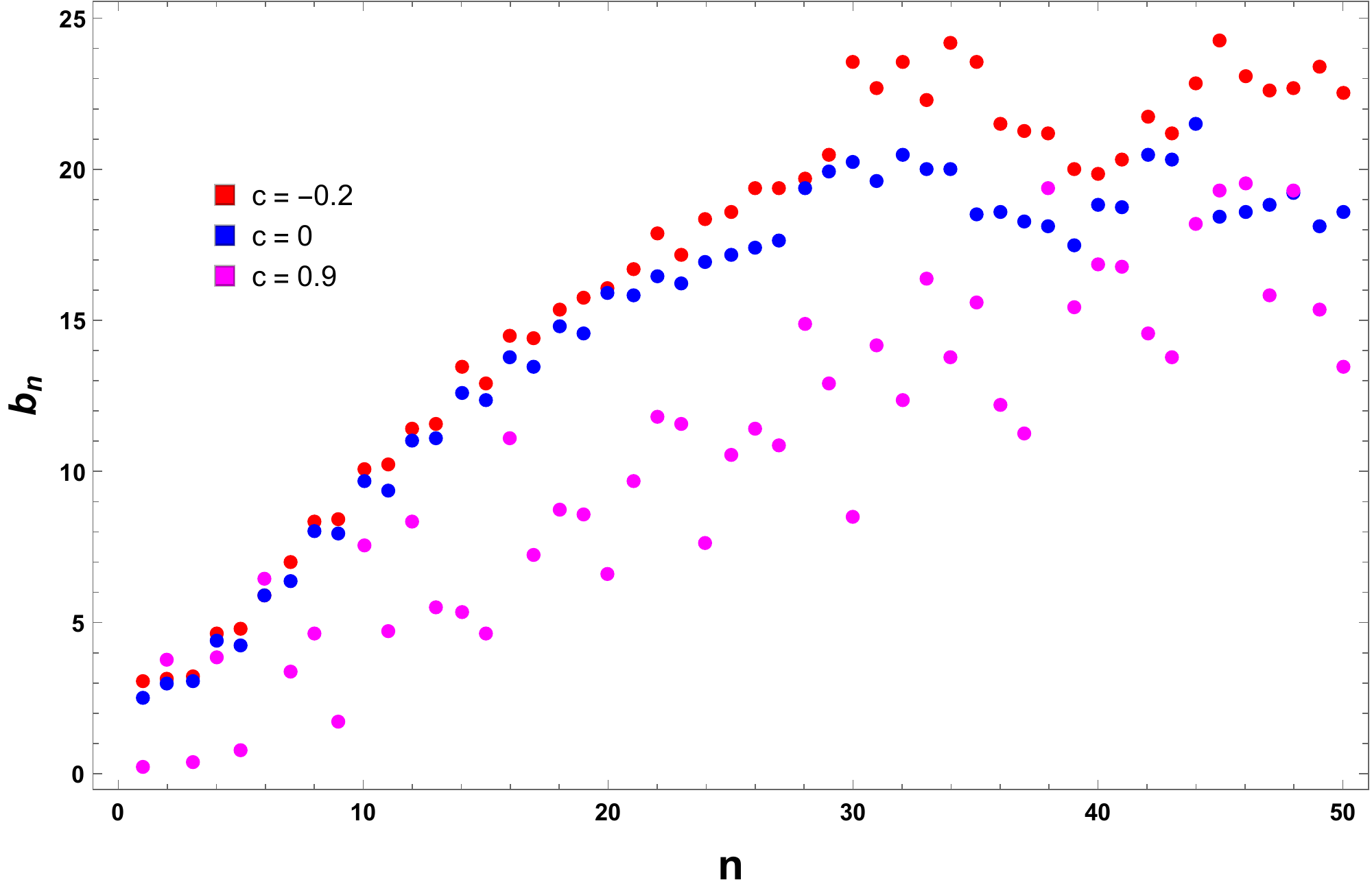}
			\caption{Growth of $b_n$ with $n$.}
			\label{figf1y}
		\end{subfigure}
		\caption{Saddle-point analysis of the classical FP model. (a) shows the variation of the exponent with parameter $c$. The largest growth is obtained for $c = -0.2$. The slope of the linear growth of the Lanczos coefficients ($\alpha$) for different values of $c$ is also plotted for $s_{1} = s_{2} \equiv s = 75$ and for the operator choice $\hat{O} = \hat{x}_{1} + \hat{x}_{2}$. As is evident, the bound is saturated only for $c = 0$. Away from $c = 0$, we always have $2\alpha > \omega(c)$. %\color{red}{Legend is wrong: should be $\alpha$ and $\omega(c)/2$} 
		(b) The Lanczos coefficients $b_n$ for three different values of $c$.}
		\label{figomegac}
\end{figure}

With this, we can simplify \eqref{fpsad} to obtain the following saddle points
\begin{align}
    (x_{1},y_{1},z_{1})&=(x_{2},y_{2},z_{2}) = (\pm 1,0,0)\,, \label{fpsad1} \\
    (x_{1},y_{1},z_{1}) &= \Bigg(\frac{(1+c)\gamma}{4(1-c)}, \,0\, ,\sqrt{1 - \bigg(\frac{(1+c)\gamma}{4(1-c)}\bigg)^{2}} \Bigg)\,, \nonumber \\
    (x_{2},y_{2},z_{2}) &= \Bigg(\frac{(1+c)}{4(1-c)\gamma}, \,0 \,,\sqrt{1 - \bigg(\frac{(1+c)}{4(1-c)\gamma}\bigg)^{2}} \Bigg)\,. \label{fpsad2}
\end{align}
where $ \frac{1+c}{4(1-c)}\leq \gamma \leq \frac{4(1-c)}{1+c}$. \par 
The Jacobian matrix for this time evolution can be written as
\begin{equation}
\mathcal{J} = \begin{bmatrix}
    0 & -4 (1-c)z_{2} & 0 & 0 & 0 & -4(1-c)y_{1}\\
    4 (1-c) z_{2} & 0 & -(1+c) & 0 & 0 & 4(1-c)x_{1}\\
    0 & 1+c & 0 & 0 & 0 & 0 \\
    0 & 0 & -4(1-c)y_{2} & 0 & -4(1-c)z_{1} & 0 \\
    0 & 0 & 4(1-c)x_{2} & 4(1-c)z_{1} & 0 & -(1+c) \\
    0 & 0 & 0 & 0 & 1+c & 0
    \end{bmatrix}\,.
\end{equation}
Note that \eqref{fpsad2} suggests that there are an infinite number of such saddle points, corresponding to every value of $\gamma$ for a given $c$. One can see numerically, that the eigenvalues of $\mathcal{J}$ for such points are all complex. Therefore, these are not unstable saddle points. \par
We investigate only the saddle given by \eqref{fpsad1}, which we can plug into $\mathcal{J}$. Thus, we get a matrix with the following eigenvalues
\begin{align}
    \{ \omega(c), 0, -\omega(c), i \omega(c), 0, - i \omega(c) \}\,,
\end{align}
where $\omega(c) = \sqrt{(1+c)(3-5c)}$, for $c \in [-1, 3/5]$ (and vanishes elsewhere). Therefore \eqref{fpsad1} is an unstable saddle point with an unstable exponent $\omega(c)$ (i.e., $\lambda_{\text{saddle}}$). The maxima is obtained at $c = -0.2$, where $\omega(c) = 4/\sqrt{5}$ (Fig.\,\ref{fig:omega}). 
Extensive numerical calculation of the OTOC~\cite{Xu:2019lhc} indicates that the scrambling in the FP model is dominated by this saddle: the bound $\lambda_L \ge \omega(c)$ appears to be saturated. We can equivalently compute the growth of the Lanczos coefficients, which is shown in Fig.\,\ref{figf1y} for three different values of $c$.

\begin{figure}[t]
		\centering
		\includegraphics[width=0.75\textwidth]{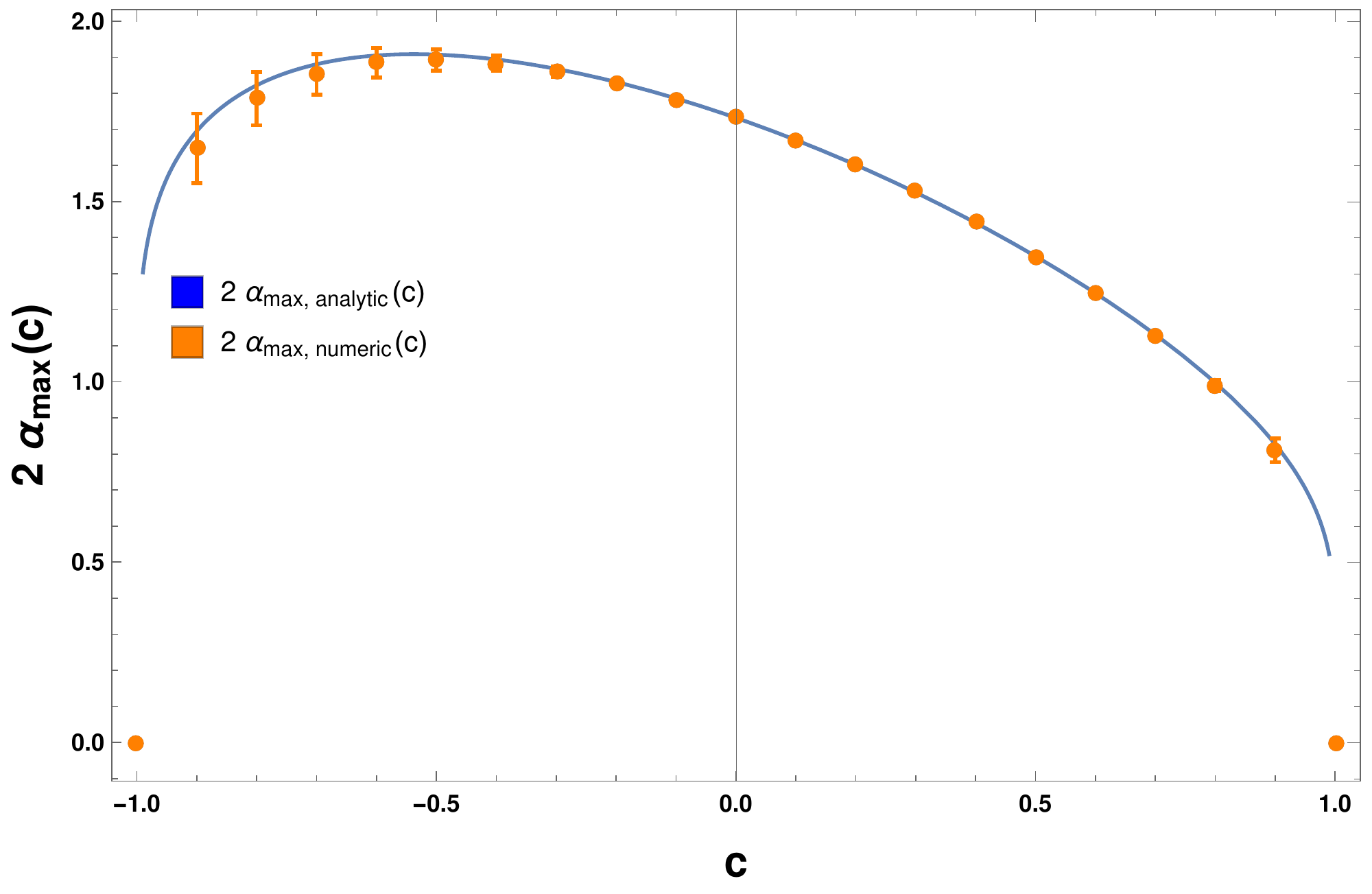}
		\caption{Evaluation on the semi-analytic bound on $2\alpha$. The analytic curve is obtained by maximizing the value of $2\alpha(E)$ over the allowed energy range and finding the extremum for each value of $c$. The result closely matches  (especially near the maximally chaotic region) with the slope obtained from linear growth of Lanczos coefficients for the FP model in the classical limit (i.e., $S \rightarrow \infty$). For this purpose, we take the first $50$ Lanczos coefficients to determine the slope of the linear growth.
		}
		\label{fig:tightfit}
\end{figure}

\subsection{Lanczos coefficients and a tight bound on $\alpha$}

The K-complexity growth rate $2 \alpha$ is also bounded below by $\omega(c)$ (since $2\alpha \ge \lambda_L$), yet this bound is not tight. This was shown in Ref.\,\cite{Parker:2018yvk}. We replicated this numerical observation by performing the Lanczos algorithm both in classical ($S\to\infty$) limit and with quantum systems, see Fig.\,\ref{fig:omega} (the Lanczos coefficients can be evaluated by employing the Lanczos algorithm with the commutators replaced by Poisson brackets). The bound $2 \alpha \ge \lambda_L$ appears only tight at $c = 0$.

Now, we propose another below bound on $\alpha$, which we conjecture to be tight. For this, we consider the equal-spin, subspace of the FP phase space, i.e., defined by $s_1 = s_2 = s$. It is straightforward to see that the FP dynamics preserves this subspace, and reduces to one described a re-scaled LMG Hamiltonian: 
\begin{align}
    H = (1+c)(x_1 + x_2)+ 4(1-c) z_{1}z_{2} \,
      \xrightarrow[]{s_{1}=s_{2}=s} \,2(1+c)\left(x +  \frac{2(1-c)}{1+c} z^2\right)\,.
\end{align}
In terms of the LMG coupling constant, we have $J= 2 (1-c)/(1+c)$. In particular, $c = 1,0,-1$ corresponds to $J = 0, 2, \infty$, respectively. We then expect the inequality:
\begin{align}\label{eq:bound_alpha_FP}
    \alpha_{\text{FP}} \geq 2(1+c) \, \alpha_{\text{LMG}}\left( J = \frac{2(1-c)}{1+c} \right)\,.
\end{align}
This is because the K-complexity growth rate of the infinite-temperature ensemble should be no smaller than any sub-ensemble, including the equal-spin one. The RHS of \eqref{eq:bound_alpha_FP} involves is in turn a maximum over energies, $\alpha_{\text{LMG}}= \sup_{E} \alpha_{\text{LMG}}(E)$, and can be thus calculated semi-analytically using \eqref{eqt2}. The result is plotted in Fig.~\ref{fig:tightfit} and compared to the numerical estimates of $\alpha_{\text{FP}}$. We observe that the bound \eqref{eq:bound_alpha_FP} is saturated (within error bars) throughout the interval $c\in (-1, 1)$. This is a surprising result since it suggests the K-complexity growth of the FP model is dominated by the equal-spin sub-manifold, in which the dynamics is not chaotic. It also naturally explains why the saddle-point bound $\alpha \ge \omega(c)/2$ is only tight at $c=0$: it corresponds to $J=2$, the only value where the same bound is saturated in the LMG model.

\bibliographystyle{JHEP}
\bibliography{references}  

\end{document}